\documentclass[openacc]{rsproca_new}

\newtheorem{theorem}{\bf Theorem}[section]

\usepackage{subfigure}
\usepackage{graphicx}
\graphicspath{{Figures/}}
\usepackage[all]{xypic}
\usepackage{bm}

\usepackage{amsmath,amsfonts,amssymb}
\newcommand{\R}{\mathbb{R}} 
\newcommand{\N}{\mathbb{N}}
\newcommand{\Hc}{\mathcal{H}}

\newcommand{\be}{\begin{equation}}
\newcommand{\ee}{\end{equation}}

\begin{document}
\title{Simple, low-cost, and accurate data-driven geophysical forecasting with learned kernels}

\author{
B. Hamzi$^{1}$, R. Maulik$^{2}$ and H. Owhadi$^{3}$}

\address{$^{1}$Department of Mathematics, Imperial College London, London, SW7 2AZ, UK.\\
$^{2}$Argonne Leadership Computing Facility,
Argonne National Laboratory, Lemont, IL 60439, USA.\\
$^{3}$Department of Computing and Mathematical Sciences, Caltech,  Pasadena, CA 91125, USA.}

\subject{Atmospheric Science, Artificial intelligence, Mathematical modeling}


\corres{Romit Maulik\\
\email{rmaulik@anl.gov}}

\begin{abstract}
Modeling geophysical processes as low-dimensional dynamical systems and regressing their vector field from data is a promising approach for learning emulators of such systems. We show that when the kernel of these emulators is also learned from data (using kernel flows, a variant of cross-validation), then the resulting data-driven models are not only faster than equation-based models but are easier to train than neural networks such as the long short-term memory neural network. In addition, they are also more accurate and predictive than the latter. When trained on geophysical observational data, for example, the weekly averaged global sea-surface temperature, considerable gains are also observed by the proposed technique in comparison to classical partial differential equation-based models in terms of forecast computational cost and accuracy. When trained on publicly available re-analysis data for the daily temperature of the North-American continent, we see significant improvements over classical baselines such as climatology and persistence-based forecast techniques. Although our experiments concern specific examples, the proposed approach is general, and our results support the viability of kernel methods (with learned kernels) for interpretable and computationally efficient geophysical forecasting for a large diversity of processes.
\end{abstract}


\begin{fmtext}

\end{fmtext}


\maketitle

\section{Introduction}

The numerical simulation of geophysical processes for forecasting is limited by high computational costs, complex parameterization requirements, and the spatio-temporal resolution of fast processes. Data-driven methods are becoming increasingly popular for surrogate modeling (emulation) or data analyses of all or a portion of the processes involved in numerical weather prediction \cite{o2018using,huntingford2019machine,rolnick2019tackling,schmidt2019spectral}. These include deep learning and neural network methods \cite{van2000neural,toms2019deep,scher2018toward,prabhat2017climatenet}, which have shown exceptional performance at the cost of a drop in interpretability.  As an alternative, we combine interpretable kernel methods with kernel selection methods (Kernel Flows \cite{Owhadi19}, as presented in \cite{BH_HO_KFs} for learning dynamical systems) for the {forecasting of geophysical processes using data} and show competitive results at exceptionally low computational costs for sub-seasonal temperature forecasting when the underlying kernels are also learned from data. In particular, we demonstrate that these methods provide interpretable and computationally efficient alternatives\footnote{Trained neural networks themselves can also be viewed as kernel machines \cite{pedro, montavon} or warping kernels regressors \cite{owhadi2020ideas}. } to deep learning architectures, which are infrastructure hungry and require off-nominal post-hoc analyses for model interpretability such as Shapley additive explanations \cite{lundberg2017unified} or layerwise relevance propagation \cite{bach2015pixel}. The framework proposed here is amenable to complex dynamics emulation, a-posteriori error estimation, {and input and model-form uncertainty-quantification while being computationally inexpensive from a learning perspective}. However, we emphasize that the goal of this study is \emph{not} to claim superiority over other geophysical emulators but to advocate the use of simple data-driven methods as effective baselines before migrating to compute hungry and poorly interpretable deep learning methods.

{The rest of this document is organized as follows: Section 2 contains a literature review summarizing recent work in data-driven geophysical forecasting. Section 3 introduces the specific data sets that have been used for forecasting. The proposed algorithms for forecasting are introduced in Section 4. Section 5 outlines results from experiments and a discussion with concluding remarks is provided in Section 6. }

\section{Related work}

Temperature forecasting is a crucial capability for several applications relevant to agriculture, energy, industry, tourism, and the environment. Improved accuracy in short and long-term forecasting of air and sea-surface temperature has significant implications for cost-effective energy policy, infrastructure development, and downstream economic consequences \cite{dell2012temperature}. The current state of the art in temperature forecasting is obtained with partial differential equation (PDE) based methods \cite{saha2010ncep,molteni1996ecmwf}. Since these methods generally require solving large systems of equations with high-performance computing resources, they are limited by access and considerations of energy-efficiency. 

Therefore, the task of temperature (and more broadly, geophysical process) forecasting has recently become a popular application of machine learning (ML) methods due to the promise of comparable (if not greater) forecast accuracy at a fraction of the computational cost. This also enables uncertainty quantification through ensemble forecasts \cite{lakshminarayanan2016simple}, which are impossible for PDE-based methods due to their excessive computational complexity. For these reasons, there has been a great degree of interest in building ML `emulators' or `surrogate models' from various geophysical data sets. There have been several studies on the use of machine learning for accelerating geophysical forecasts in recent times. Several rely on using machine learning methods to devise parameterizations for processes that contribute a high cost to the numerical simulation of the weather and climate \cite{chattopadhyay2020data,gentine2018could,brenowitz2018prognostic,rasp2018deep,bolton2019applications,zanna2020data}. In such cases, partial differential equations are not eschewed entirely. Other studies have looked at complete system emulators (i.e., forecasting from data alone) with a view to forecasting without any use of (and consequent limitations of) equation-based methods \cite{liu2016application,scher2018toward,nooteboom2018using,weyn2020improving,rasp2020purely,chattopadhyay2020analog,rodrigues2018deepdownscale}. Other studies have looked at utilizing historical information for data-driven forecasting of specific processes \cite{shi2015convolutional,maulik2020recurrent,skinner2020meta,dueben2018challenges} by focusing on specific influential variables or through the use of time-delay embeddings to offset inaccuracies due to unresolved variables. Further opportunities and perspectives for the use of data-driven methods for geosciences may be found in the reviews of \cite{karpatne2018machine,dueben2018challenges}. {Before proceeding, we note that a vast majority of the data-driven developments for geophysical forecasting have involved the use of variants of deep learning methods, for example ResNets\cite{rasp2021data}, CapsuleNets \cite{chattopadhyay2020data}, U-Nets \cite{weyn2020improving}, LSTMs \cite{maulik2020recurrent}, Convolutional-LSTMs \cite{shi2015convolutional}, neural ordinary differential equations \cite{portwood2019turbulence,maulik2020time}, local or global fully connected deep neural networks \cite{dueben2018challenges}. These methods, while exceptionally powerful in learning complex functions, hamper interpretability and require large computational resources for optimization. Furthermore, extensions to model-form and data uncertainty quantification aggravate computational requirements significantly. Therefore, the goal of this research is to put forth viable alternatives to deep learning-based geophysical forecasting via the use of gray-box kernel flows for learning dynamical systems. }

{In this article, we introduce an entirely data-driven method for forecasting the weekly-averaged sea-surface temperature for the entire planet and the daily maximum air temperature over the North-American continent. Furthermore, our method is developed to provide forecasts without the requirement of large-scale computational resources and with a greater degree of transparency}. We achieve this by obtaining a low-dimensional affine subspace approximation of the temperature field on which a reduced system is evolved. Both dimensionality reduction and system evolution are performed using data-driven techniques alone, with the former employing a proper-orthogonal decomposition and the latter combining kernel methods with a cross-validation technique known as Kernel Flows. We remark that in contrast to the growing popularity of deep learning methods for forecasting, we propose using classical dimensionality reduction and time-series forecasting with suitable inductive biases. {Here, inductive biases are modifications to machine learning frameworks that ``assist" learning given prior knowledge of the physics they are employed to learn.} Our competitive results motivate the creation of simple data-driven baselines that compare favorably to PDE-based methods without specialized neural architectures. At this point, we would like to emphasize to the reader that while the experiments in this article devote themselves to forecasts on temperature, the proposed framework is general and can be applied to other geophysical flow-field forecasting, provided linear or non-linear low-dimensional embeddings may be extracted, on which the dynamics evolve.

\section{Data set(s)}

\subsection{Weekly averaged sea surface temperature}

For our first experiment, we use the open-source National Oceanic and Atmospheric Administration (NOAA) Optimum Interpolation sea surface temperature V2 data set (henceforth NOAA-SST).\footnote{Available at https://www.esrl.noaa.gov/psd/}  This data set has a strong periodic structure due to seasonal fluctuations in addition to rich fine-scaled phenomena due to complex ocean dynamics. Weekly-averaged NOAA-SST snapshots are available on a quarter-degree grid which is sub-sampled to a one-degree grid for the purpose of demonstrating our proposed methodology in a computationally efficient manner. This data set, at the 1-degree resolution, has previously been used in several data-driven analysis tasks (for instance, see \cite{kutz2016multiresolution,callaham2019robust} for specific examples), particularly from the point of view of extracting seasonal and long-term trends as well as for flow-field recovery \cite{maulik2020probabilistic}. Each ``snapshot'' of data originally corresponds to an array of size 360 $\times$ 180 (i.e., arranged according to the longitudes and latitudes of a one-degree resolution). However, for effective utilization in forecasting, a mask is used to remove missing locations in the array that corresponds to the land area. Furthermore, it should be noted that forecasts are performed for those coordinates which correspond to oceanic regions alone, and inland bodies of water are ignored. The non-zero data points then are subsequently flattened to \textcolor{black}{obtain a column vector for each snapshot of our training and test data.}

This data is available from October 22, 1981, to June 30, 2018 (i.e., 1,914 snapshots for the weekly averaged temperature). We utilize the period of October 22, 1981, to December 31, 1989. The rest (i.e., 1990 to 2018) is used for testing. Our final number of snapshots for training amounts to 427, and for testing amounts to 1487. This train-test split of the data set is a common configuration for data-driven studies \cite{callaham2019robust} and the 8-year training period captures several short and long-term trends in the global sea surface temperature. Individual training samples are constructed by selecting a window of inputs (from the past) and a corresponding window of outputs (for the forecast task in the future) from the set of 427 training snapshots. \textcolor{black}{From the perspective of notation, if $\bm{\theta}_t$ is a snapshot of training data, $t= 1,2,\dots,427$, a forecasting technique may be devised by learning to predict $\bm{\theta}_{t+1}, \dots, \bm{\theta}_{t+\tau}$ given $\bm{\theta}_t, \bm{\theta}_{t-1}, \dots, \bm{\theta}_{t-\tau}$. We note that this forecast is performed non-autoregressively---that is, the data-driven method \emph{is not} utilized for predictions beyond the desired window size $\tau$. Therefore, it is always assumed that the \emph{true} $\bm{\theta}_t, \bm{\theta}_{t-1}, \dots, \bm{\theta}_{t-\tau}$ is available prior to making predictions.} This means that given a window of true inputs (for example obtained via an observation of the state of the system given re-analysis data), a forecast is made for a series of outputs (corresponding to the window length) before a metric of accuracy is computed for optimization. This is in contrast to a situation where simply one step of a prediction is utilized for computing a fitness metric which adversely affects the ability of the predictive method for longer forecasts into the future (due to amplifying errors with each forecast step). The window size for the set of experiments on this data set is fixed at 8 weeks. The window-in and window-out construction of the training data leads to a final training data set of size 411 samples. Since this data set is produced by combining local and satellite temperature observations, it represents an attractive forecasting task for \emph{non-intrusive} data-driven methods without requiring the physical modeling of underlying processes.

\subsection{North-American daily midnight surface temperature}

The National Centers for Environmental Prediction's (NCEP) North-American Mesoscale Forecast System (NAM) \cite{saha2010ncep} is one of the main mesoscale models used for guiding public and private sector meteorologists. NAM runs four times daily at three different spatial scales: (1) Full North-American 12-km resolution; (2) 4-km Continental U.S. (CONUS) nest, 6-km Alaska nest, and 3-km Hawaii and Puerto Rico nests. These domains are one-way nested inside the 12-km domain; (3) High-resolution nested domain which has a different location each cycle based upon the NCEP Service Centers and National Weather Service Offices. For this work, analysis data from NAM was used, using the 12-km resolution grid for the surface temperature. The NAM data is collected in a time period between the 28th of October 2008 to the 20th of September 2018 on a daily cadence. In particular, we measure the temperature at midnight on each day in this temporal domain. This corresponds to 3569 snapshots of sea and land surface temperature data. In contrast to the previous problem, our goal is to forecast one week in advance with the temperature resolved daily. The first 2555 snapshots are reserved for the purpose of training the proposed forecasting technique, while the rest are used for testing. In a manner similar to the previous test case, a time delay of 7 days is used to specify the inputs to obtain the 7-day forecast of temperature in the future.

\section{Methods}

\begin{figure}
    \centering
    \includegraphics[width=\textwidth]{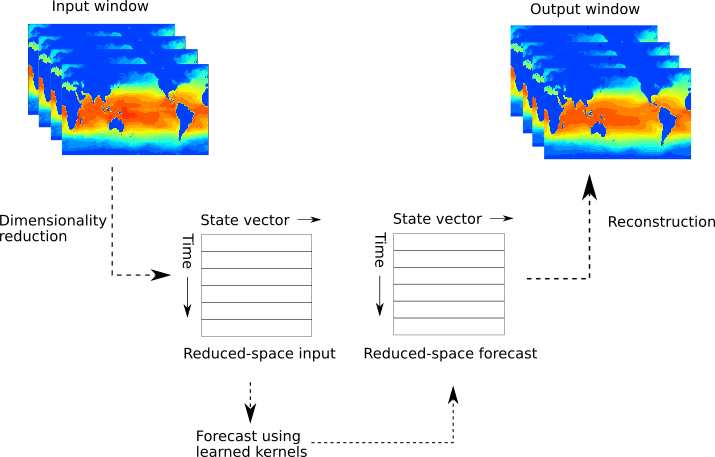}
    \caption{The overall schematic of the proposed workflow. Learned kernels are used to forecast the state of a geophysical process on a low-dimensional embedding obtained by the proper orthogonal decomposition. Forecasted trajectories in the reduced subspace are reconstructed to obtain the final predicted state.}
    \label{RKHS_Schematic}
\end{figure}

{In this section, we shall introduce our proposed algorithm for forecasting. First, we seek to reduce the degrees of freedom of our forecast problem by performing a dimensionality reduction using the proper orthogonal decomposition (POD) . This reduced representation of our dataset is then used to train an efficient dynamical systems emulator using kernel methods. Forecasts from this emulator may then be reconstructed in the original space through the use of previously computed POD basis functions. The overall schematic of this procedure is shown in Figure \ref{RKHS_Schematic}.}

\subsection{Dimensionality reduction: Proper orthogonal decomposition}

POD provides a systematic method to project dynamics of a high-dimensional system onto a lower-dimensional subspace. We suppose that a single snapshot of the full system is a vector in $\mathbb{R}^N$, where $N$ could be the number of grid points at which the field is resolved. Observing the system across a number of time points gives us the snapshots $\bm{\theta}_1, \dots, \bm{\theta}_T$, with mean subtracted by convention. The aim of POD is to find a small set of orthonormal basis vectors $\bm{v_1}, \dots, \bm{v}_M$, with $M \ll N$, which approximates the spatial snapshots,
\be
\bm{\theta}_t \approx \sum_{j=1}^M a_j(t)\bm{v}_j, \quad t=1,\dots,T,
\ee
and so allows us to approximate the evolution of the full $N$ dimensional system by considering only the evolution of the $M$ coefficients $a_j(t)$. POD chooses the basis, $\bm{v}_j$, to minimize the residual with respect to the $L_2$ norm, 
\be
R = \sum_{t=1}^{T} || \bm{\theta}_t - \sum_{j=1}^M a_j(t) \bm{v}_j ||^2.
\ee
Defining the snapshot matrix, $\bm{S} = [ \bm{\theta}_1 | \cdots | \bm{\theta}_T ]$, the optimal basis is given by the $M$ eigenvectors of $\bm{SS}^T$, with largest eigenvalues, after which, the coefficients are found by orthogonal projection, $\mathbf{a}(t) = \langle \bm{\theta}_t,\bm{v}\rangle$~\cite{ModalAnalysisRev}. {The coefficients $\mathbf{a}(t)$ correspond to a time-series with an $M-$dimensional state vector which is the focus of our temporal forecasting.}

For both of our data sets, we take only the training data snapshots, say $\bm{D}_1, \dots,\bm{D}_T$, from which we calculate the mean $\bar{\bm{D}} = (1/T) \sum_t \bm{D}_t$, hence defining the mean subtracted snapshots $\bm{\theta}_t = \bm{D}_t - \bar{\bm{D}}$. We then create the snapshot matrix, $\bm{S}$, and find numerically the $M$ eigenvectors of $\bm{S}\bm{S}^T$ with largest eigenvalues. {From this, we train models, $f^\dagger$ (to be defined in the next section), to forecast the coefficients
\textcolor{black}{
\begin{align}
(\bm{a}(t+1), \bm{a}(t+2), \dots, \bm{a}(t+\tau)) & \approx (\hat{\bm{a}}(t+1), \hat{\bm{a}}(t+2), \dots, \hat{\bm{a}}(t+\tau)) \\
& = f^\dagger ( \bm{a}(t), \bm{a}(t-1),\dots, \bm{a}(t-\tau)). \nonumber
\end{align}
}
making predictions of future coefficients given previous ones. Here, $\tau$ corresponds to the time-delay that is embedded into the input feature space for the purpose of forecasting. In this article, $\tau$ also stands for the forecast length obtained from the fit model although in the general case, these quantities can be chosen to be different.} \textcolor{black}{We can now test for predictions on unseen data, $\bm{E}_1,\dots,\bm{E}_Q$, where $\bm{E}_t$ is an unseen snapshot of data at time $t$ and $Q$ is the total number of test snapshots. Note that these test snapshots are obtained for a time-interval that has not been utilized for constructing the POD basis vectors.} We proceed by taking the mean $\bar{\bm{D}}$, and vectors $\bm{v}_j$ calculated from the training data to get test coefficients at an instant in time
\be
a_j(t) = \langle \bm{E}_t - \bar{\bm{D}}, \bm{v}_j \rangle, \quad j = 1,\dots,M,
\ee
which will be used with the model $f^\dagger$ to make future predictions in the unseen testing interval. The prediction for the coefficients $\hat{\bm{a}}$, can be converted into predictions in the physical space by taking $\bar{\bm{D}} + \sum_j \hat{a}_j \bm{v}_j$. This procedure only makes use of testing data to pass into the model, not to train the model in any way. {Crucially, to make a forecast of $\bm{E}_{t+1}, \bm{E}_{t+2}, \dots, \bm{E}_{t+\tau}$, only previous measurements $\bm{E}_{t},\bm{E}_{t-1}, \dots, \bm{E}_{t-\tau}$ are needed.}

\subsection{Cross-validation for temporal forecasting}

The method employed for temporal forecasting was introduced in \cite{BH_HO_KFs} for learning dynamical systems from data. It is a { kernel  regression} with a kernel learned from data from a variant of cross-validation known as Kernel Flows \cite{Owhadi19}.

To describe this method consider the problem of  forecasting $\bm{a}(n+1)$ given the observation of $\bm{a}(1),\ldots,\bm{a}(n)$, where  $\bm{a}(1),\ldots,\bm{a}(k),\ldots$ is a time series in $\R^M$ (in our application, $M$ is the number of modes, used for prediction).
Evidently making a forecast requires some assumptions on the time series and the assumption that we work under is that it can be approximated  by a solution of a dynamical system of the form 
\begin{equation}\label{eqjhdbdjehbd}
z_{n+1}=f^\dagger(z_n,\ldots,z_{n-\tau^\dagger+1}),
\end{equation}
where the \textcolor{black}{time delay $\tau^\dagger \in \N^+$ (the positive natural numbers) and $f^\dagger$} is the vector field which may be unknown. Here, in our application, each $z_n$ is time-window of POD coefficients, i.e. of the form $z_n=(\bm{a}(n-\tau),\ldots,\bm{a}(n))$ where $\tau$ is the length of the {input and forecast windows ($\tau=8,7$ in our respective data sets).} A simple solution to our extrapolation problem is then, given \textcolor{black}{$\tau \in \N^+$}, to learn/approximate $f^\dagger$ from data (the past values of the time series), and then use the approximate vector field $f$ to define a surrogate model $$z_{n+1}=f(z_n,\ldots,z_{n-\tau+1})$$ for predicting the future states of the dynamical system. Therefore, the approximation of the dynamical system can be recast as that of regressing/interpolating $f^\dagger$ from pointwise measurements
\begin{equation}\label{eqn:fdagger}
f^\dagger(X_n)=Y_n\text{ for }n=1,\ldots, N-\tau,
\end{equation}
with $X_n:=(\bm{a}(n+\tau-1),\ldots,\bm{a}(n))$ and $Y_k:=\bm{a}(n+\tau)$. For the experiments in this study, forecasts of length $\tau$ are concatenated during testing to obtain predictions in the entire testing timescales. Given a kernel $K$, this interpolant (in the absence of measurement noise) is  
\begin{equation}\label{mean_gp}
f(x)=K(x,X) (K(X,X))^{-1} Y,
\end{equation}
where  $X=(X_{1},\ldots, X_{N-\tau})$, $Y=(Y_{1},\ldots, Y_{N-\tau})$,  $K(X,X)$ stands for the $N-\tau\times N-\tau $ matrix with entries $K(X_i,X_i)$, and $K(x,X)$ is the $N-\tau$ vector with entries $K(x,X_i)$. Writing $\xi$ for the centered Gaussian process with covariance function $K$ and $\Hc$ for the  reproducing kernel Hilbert spaces (RKHS) defined by $K$ (see  the appendix  for a reminder), recall that \eqref{mean_gp} is (1) the conditional expectation of $\xi$ given the measurements $\xi(X_i)=Y_i$, and (2) a minimax optimal approximation \cite{owhadi_scovel_2019}  of  $f^\dagger$ in $ \Hc$ (using the relative error in the reproducing kernel Hilbert space (RKHS) norm $\|\cdot\|_\Hc$ as a loss).

Evidently, the implementation of the proposed approach requires the prior selection of a kernel $K$ and here we propose (as in \cite{BH_HO_KFs}) to also learn that kernel $K$ from data using a variant of cross-validation known as Kernel Flows (KF) \cite{Owhadi19}. Given a family of kernels $K_\theta(x,x')$ parameterized by $\theta$, the KF algorithm seeks to select a $\theta$ such that subsampling the data does {not} influence the interpolant much. Writing $\|\cdot\|_{K_\theta}$ for the RKHS norm defined by the kernel $K_\theta(x,x')$, and two interpolants $u^b$ and $u^c$ obtained with the kernel $K_\theta$ and by subsampling the data ($u^c$ uses a smaller subset of the data than $u^b$), the KF algorithm seeks to learn $\theta$ by successively moving $\theta$ in the gradient descent direction of the loss $\rho=\|u^b-u^c\|^2_{K_\theta}/\|u^b\|^2_{K_\theta}$ \eqref{eqjehdhebdhdhj}.
Note that (1) $\rho$  is a cross-validation term (randomized through the subsampling procedure) acting as a surrogate for the generalization error (in relative RKHS norm), (2) $u^b$ acts as a surrogate for the target function.
To summarize, given a family of kernels $K_\theta(x,x')$ parameterized by $\theta$, the KF algorithm  is as follows \cite{Owhadi19, yoo2020deep}:
 \begin{enumerate}
     \item[i.] Randomly select subvectors $X^b$ and $Y^b$ of $X$ and $Y$ (through uniform sampling without replacement in the index set $\{1,\ldots,N-\tau\}$)
     \item[ii.] Randomly select subvectors $X^c$ and $Y^c$ of $X^b$ and $Y^b$ (by selecting, at random, uniformly and without replacement, half of the indices defining $X^b$)
     \item[iii.]  Let\footnote{ $\rho:=\|u^b-u^c\|^2_{K_\theta}/\|u^b\|^2_{K_\theta}$, with $u^b(x)=K_\theta(x,X^b) K_\theta(X^b,X^b)^{-1} Y^b$ and $u^c(x)=K_\theta(x,X^c) K_\theta(X^c,X^c)^{-1} Y^c$, and $\rho$  admits  the representation \eqref{eqjehdhebdhdhj} enabling its computation}
 \begin{equation}\label{eqjehdhebdhdhj}
 \rho(\theta,X^b,Y^b,X^c,Y^c):=1-\frac{Y^{c,T} K_\theta(X^c,X^c)^{-1} Y_c}{Y^{f,T} K_\theta(X^b,X^b)^{-1} Y^b}\,,
 \end{equation}
  be the squared relative error (in the RKHS norm $\|\cdot\|_{K_\theta}$ defined by $K_\theta$)  between
 the interpolants $u^b$ and $u^c$ obtained from the two nested subsets of the dataset and the kernel $K_\theta$
    \item[iv.] Evolve $\theta$ in the gradient descent direction of $\rho$, i.e. $\theta \leftarrow \theta - \delta \nabla_\theta \rho$
    \item[v.] Repeat until desired accuracy is achieved or computational budget is exhausted.
 \end{enumerate}

\begin{figure}
    \centering
    \includegraphics[width=\textwidth]{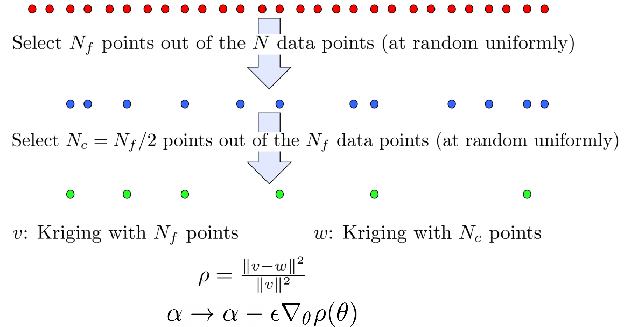}
    \caption{Learning kernel hyperparameters, $\theta$, by cross-validation from data. A workflow schematic.}
    \label{kernel_schematic}
\end{figure}

{A schematic of the kernel learning process is shown in Figure \ref{kernel_schematic}.} In this study, we use the following expression for our kernel
\begin{equation}
\begin{split}
    K_{\theta} (x,x') = 
    \theta_1^2 \exp\big(-\frac{\|x-x'\|^2}{2\theta_2^2}\big)+
     \theta_3^2 \big(1+ \theta_4^2x^T x'\big)^2
     +\theta_5^2 \big(1+\theta_6^2\|x-x'\|^2\big)^{-\frac{1}{2}}
    \\+\theta_7^2 \big(1+
     \theta_8^2\|x-x'\|^2\big)^{\theta_9}
      +\theta_{10}^2 \big(1+
     \theta_{11}^2\|x-x'\|^2\big)^{-1}
     +\theta_{12}^2\max \big(0,1-\theta_{13}*\|x-x'\|^2\big)
\end{split}
\end{equation}
{The main motivation for this kernel is the numerical experiments in \cite{BH_HO_KFs}, where the authors used kernel flows to learn prototypical chaotic dynamical systems that exhibit a rich dynamic behaviour. The motivation of the triangular kernel is that the Bernoulli map that exhibits a rich dynamic behavior is toplogically conjugate to the unit-height tent map. The quadratic kernel captures the long term correlation and the Gaussian kernel captures the short term correlation. Our experiments suggest that the particular parameterized family of kernels is not important as long as it is rich enough (to reproduce the patterns contained in the dataset).
 The additive form of the kernel is equivalent to regressing the data
 with a sum of independent Gaussian processes, each one being adapted to a particular feature/pattern of the data. As presented in
\cite{owhadi2019kernelkmd}, additive kernels can be employed to program sophisticated kernels for pattern recognition tasks.
}

In the following, we denote the use of cross-validation to forecast POD coefficients as the \emph{POD-RKHS} emulation framework.


\section{Experiments}

\begin{figure}
    \centering
    \mbox{
    \subfigure[Coefficient 1]{\includegraphics[width=0.33\textwidth]{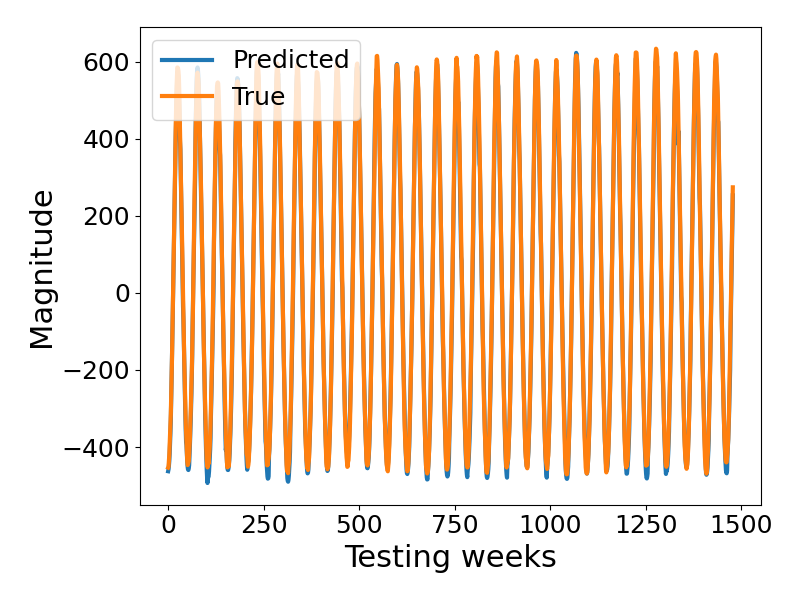}}
    \subfigure[Coefficient 2]{\includegraphics[width=0.33\textwidth]{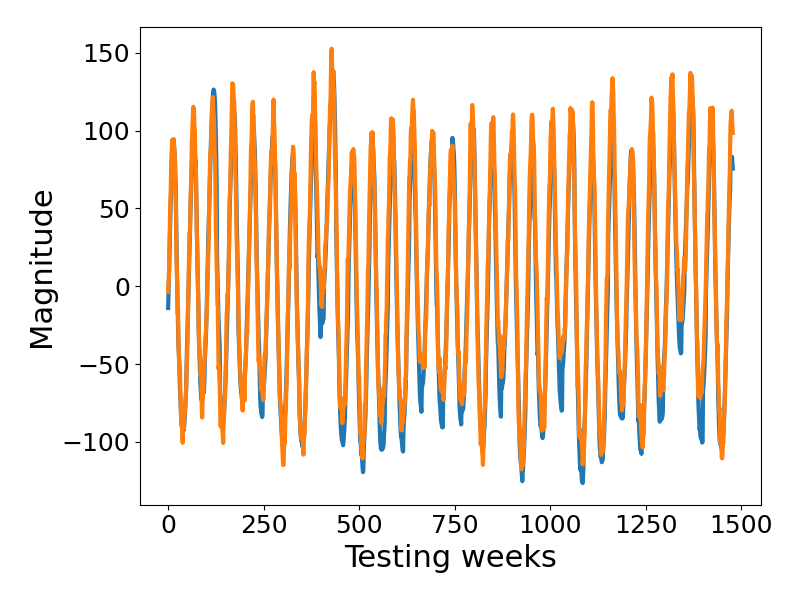}}
    \subfigure[Coefficient 3]{\includegraphics[width=0.33\textwidth]{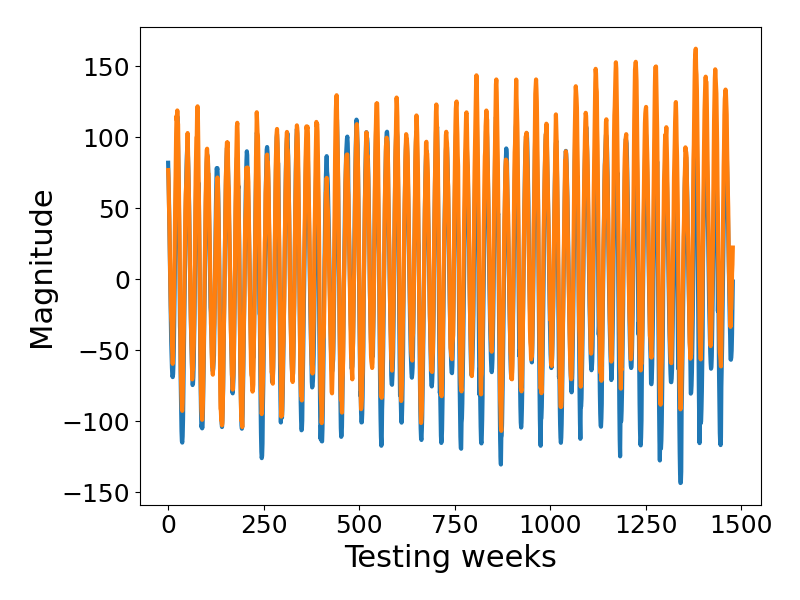}}
    } \\
    
    \mbox{
    \subfigure[Coefficient 1 Error]{\includegraphics[width=0.33\textwidth]{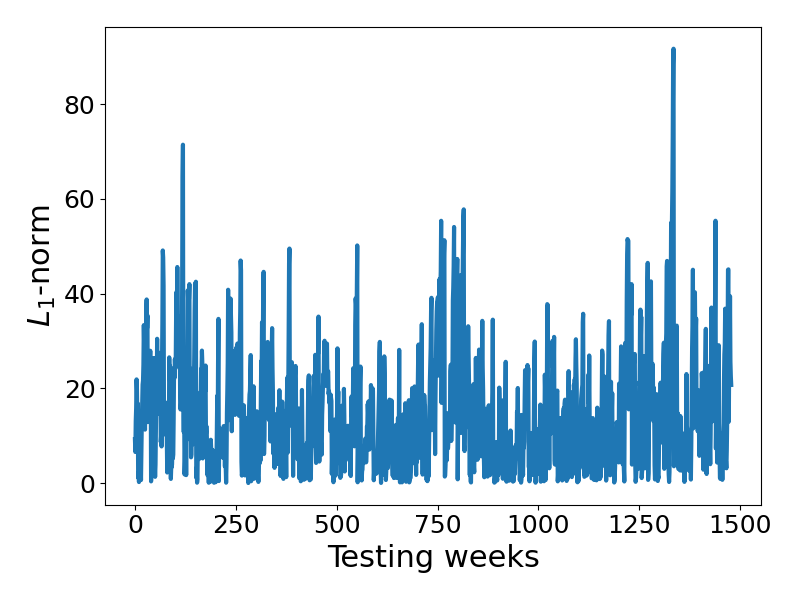}}
    \subfigure[Coefficient 2 Error]{\includegraphics[width=0.33\textwidth]{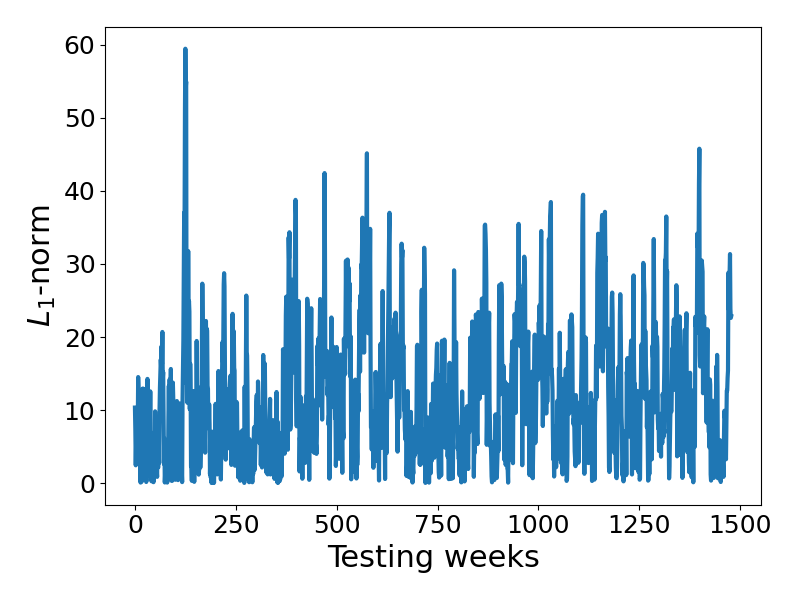}}
    \subfigure[Coefficient 3 Error]{\includegraphics[width=0.33\textwidth]{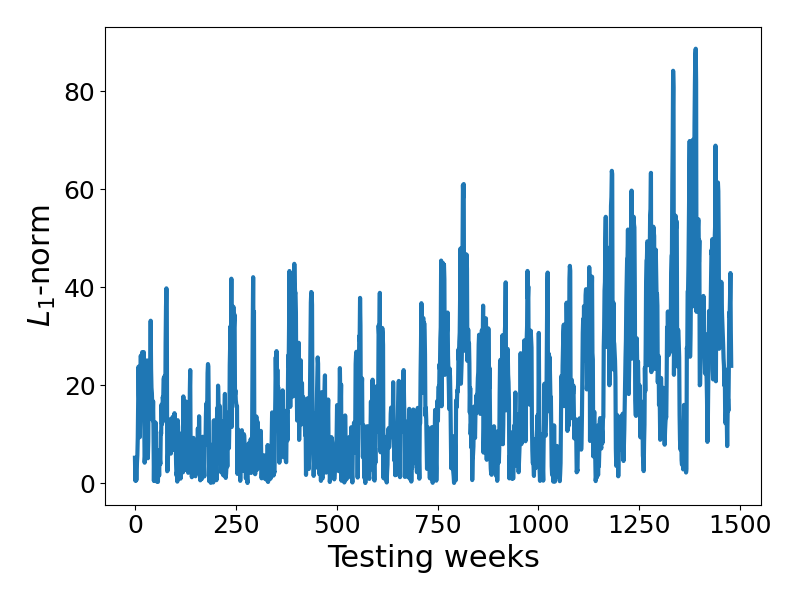}}
    }
    
    \caption{POD Coefficient predictions (top) and $L_1$ errors with corresponding true values (bottom) for the NOAA-SST data set using POD-RKHS. We utilize 427 snapshots between October 22, 1981, to December 31, 1989, for training . The remaining 1487 snapshots of the data set (i.e, 1990 to 2018) are used for testing and are shown here for model assessment.}
    \label{NOAA_Coefficients}
\end{figure}

\begin{figure}
    \centering
    \mbox{
    \subfigure[50 $^{\circ}$ latitude, 230 $^{\circ}$ longitude]{\includegraphics[width=0.48\textwidth]{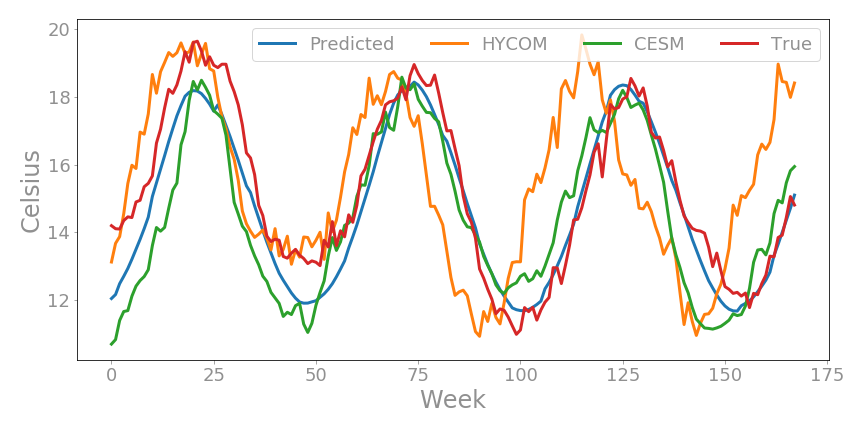}}
    \subfigure[85 $^{\circ}$ latitude, 210 $^{\circ}$ longitude]{\includegraphics[width=0.48\textwidth]{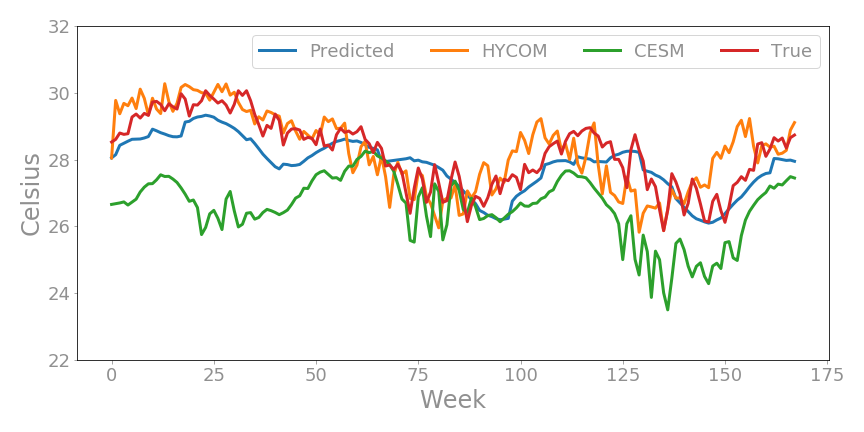}}
    }
    \caption{Probe time-series comparisons for NOAA-SST from CESM, HYCOM and POD-RKHS within the testing regime at two different locations. The data are plotted for the weeks between the 5\textsuperscript{th} of April, 2015, and the 17\textsuperscript{th} of June, 2018.}
    \label{NOAA_Probe_1}
\end{figure}

In the following section, we outline results from the use of the POD-RKHS framework for forecasting on the aforementioned data sets. We also provide metrics that evaluate the accuracy of the forecasts and compare them with baseline techniques. Our first set of results are shown for the NOAA-SST data set. The ability to forecast weekly averaged coefficients eight weeks at a time is shown in Figure \ref{NOAA_Coefficients}, (for the testing period), where it can be observed that the lower order structures are predicted with high fidelity. The framework shows deviations in the finer scale content (mode 3) as one approaches the end of the testing period indicating potential extrapolation. Similar behavior was observed in \cite{maulik2020recurrent,skinner2020meta} as well, where a \textcolor{black}{multi-cell} LSTM was used to forecast in this reduced space (henceforth POD-LSTM). Time series assessments for various point probes in the Eastern Pacific are shown for a testing sub-window {where data from all prediction sources were available (between the 5\textsuperscript{th} of April, 2015 and 17\textsuperscript{th} of June, 2018)} in Figure \ref{NOAA_Probe_1}. The plot indicates a competitive testing performance when compared to state-of-the-art equation-based methods such as the Community Earth System Model (CESM) \cite{hurrell2013community} and U.S. Navy Hybrid Coordinate Ocean Model (HYCOM) \cite{chassignet2009us}. The former is a climate simulator, and the latter is an operational forecast system giving short-term predictions. Both these techniques require vast computational resources, whereas our forecast is performed on a single node machine without any accelerator. 

\begin{figure}
    \centering
    \mbox{
    \subfigure[POD-RKHS]{\includegraphics[width=0.48\textwidth]{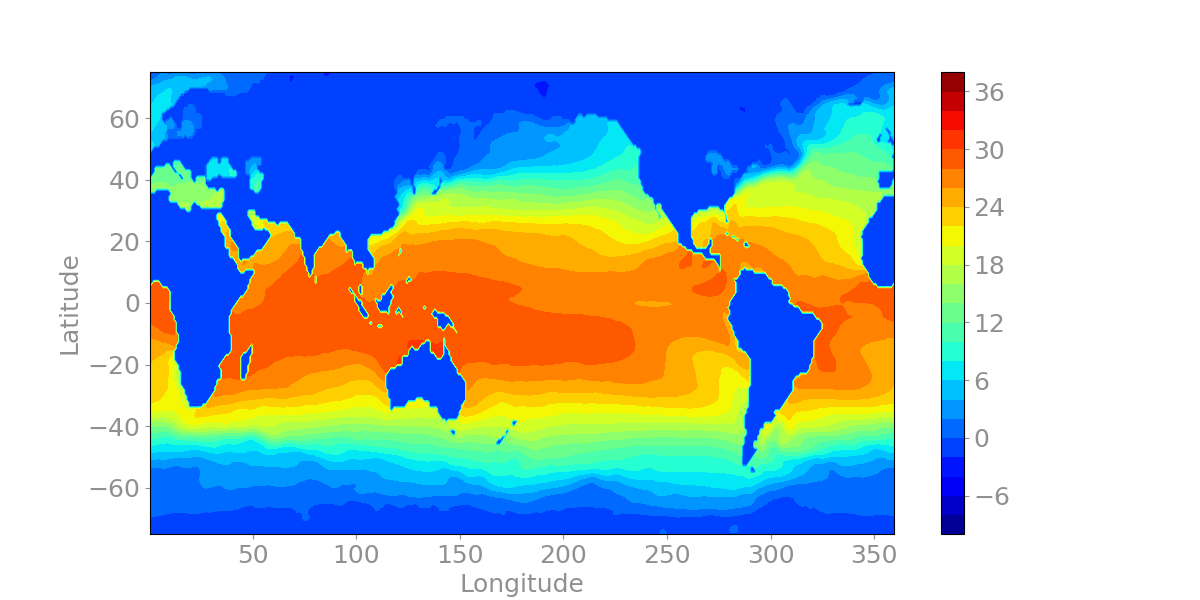}}
    \subfigure[HYCOM]{\includegraphics[width=0.48\textwidth]{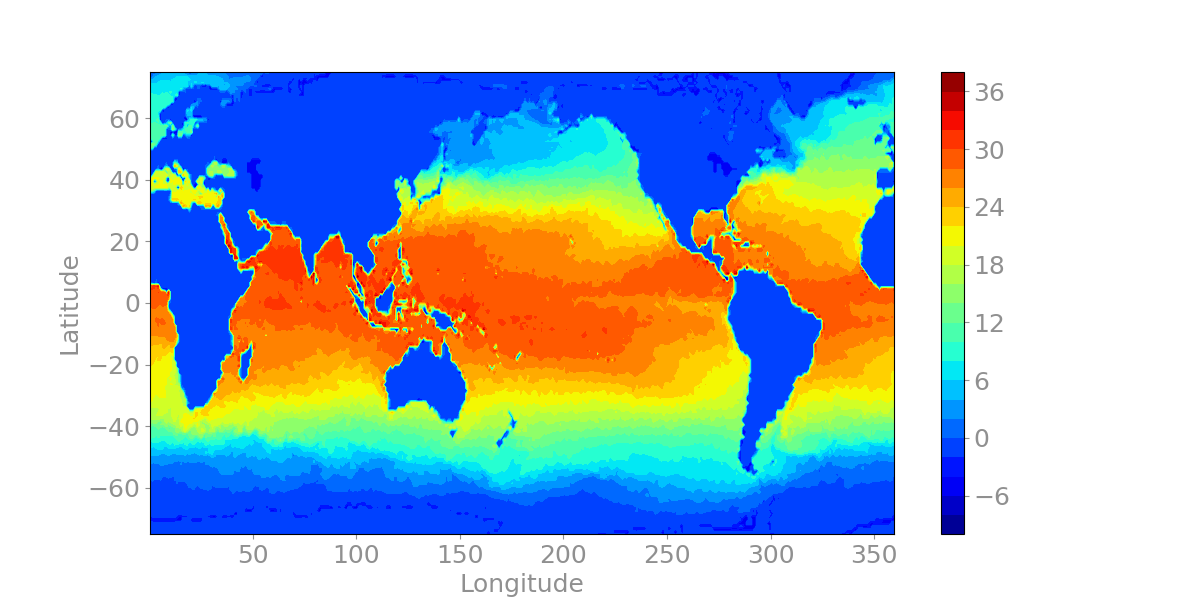}}
    } \\
    \mbox{
    \subfigure[CESM]{\includegraphics[width=0.48\textwidth]{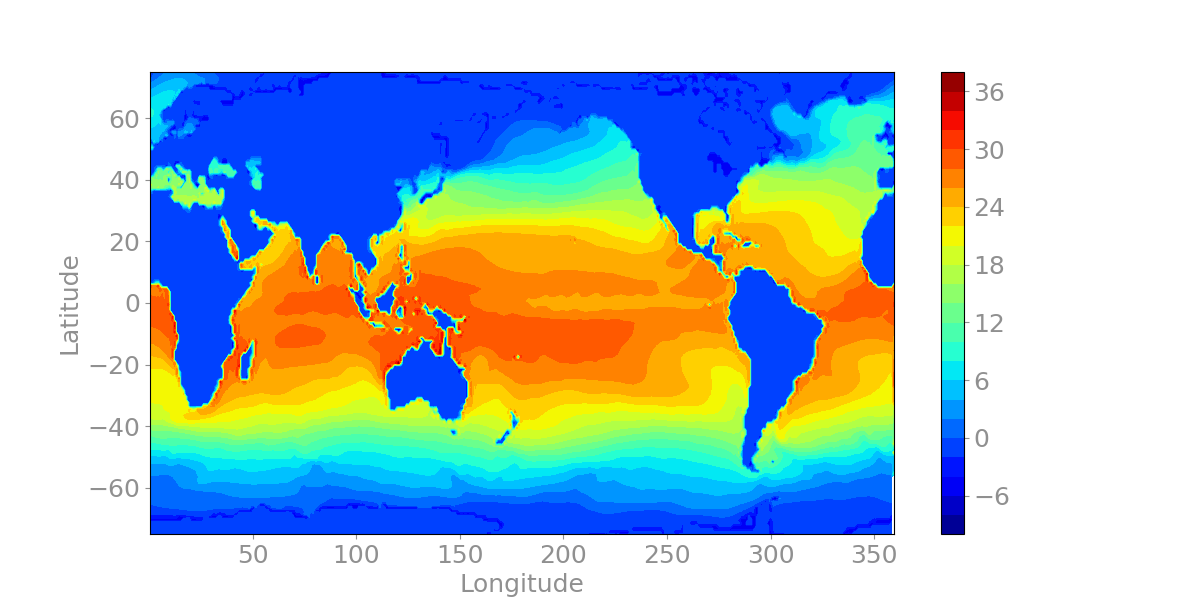}}
    \subfigure[Truth]{\includegraphics[width=0.48\textwidth]{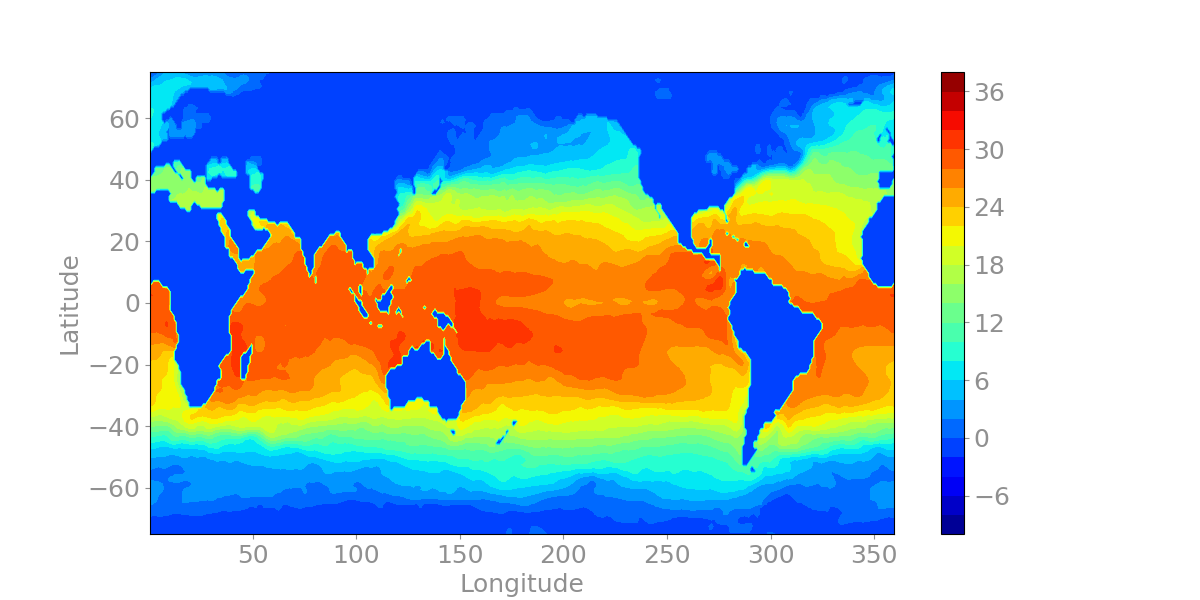}}
    }
    \caption{Contour plots for various forecasts (a-POD-RKHS, b-HYCOM, c-CESM, d-Truth), indicating how low-dimensional manifold-based emulation reduces the ability to capture fine-scaled features in the flow-field. HYCOM, with the finest resolution, is seen to capture small-scale information most accurately. Note, however, that the POD-based emulation framework is competitive in an averaged sense, as seen through RMSE metrics.}
    \label{NOAA_Contours}
\end{figure}

\begin{figure}
    \centering
    \mbox{
    \subfigure[POD-RKHS error]{\includegraphics[width=0.5\textwidth]{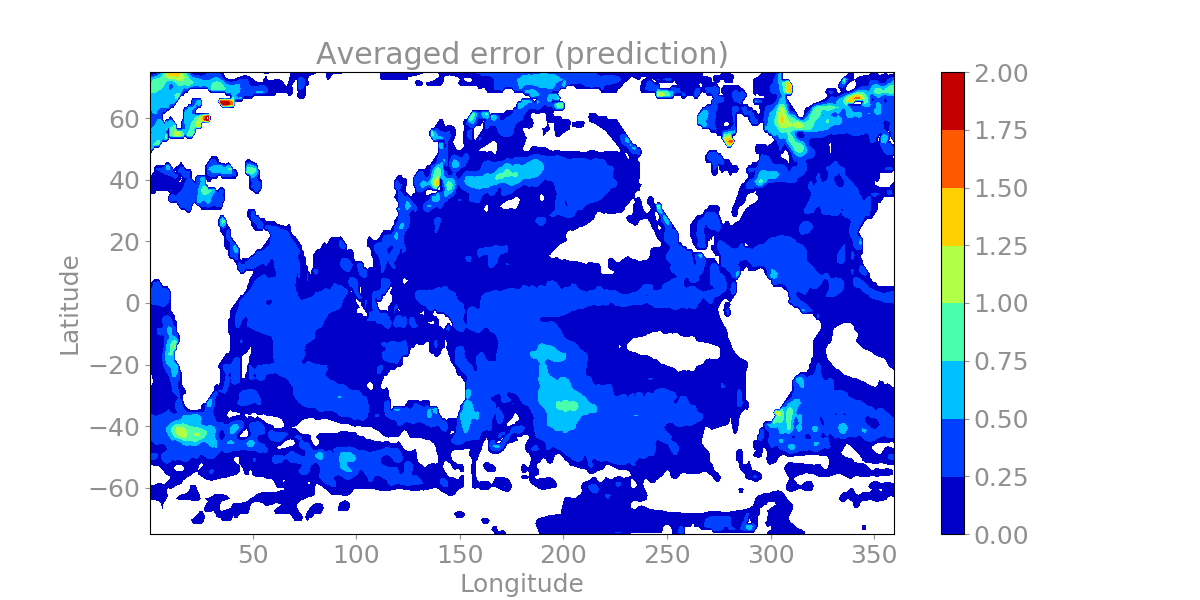}}
    \subfigure[CESM error]{\includegraphics[width=0.5\textwidth]{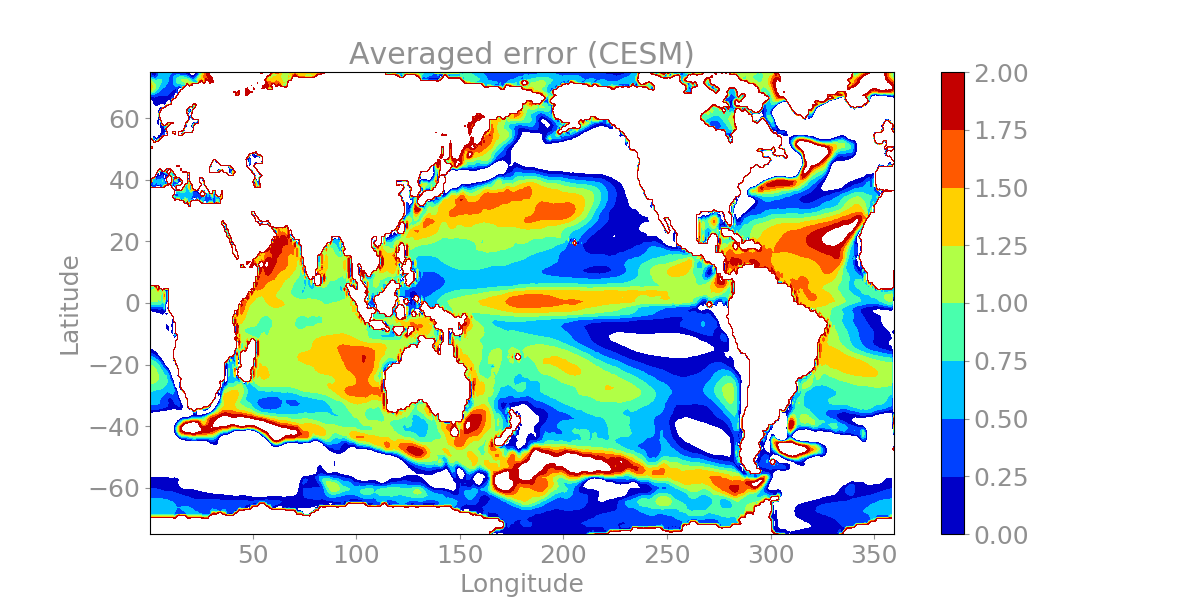}}
    } 
    \caption{Contour plots for RMSE error over the entire testing period (1990-2018) for the NOAA-SST data set for POD-RKHS (a) and CESM (b). The results indicate that the proposed low-dimensional emulator captures the long-term behavior of the SST fluctuations more accurately than CESM.}
    \label{NOAA_Contour_RMSE}
\end{figure}

Some qualitative comparisons of the predictions are shown in Figure \ref{NOAA_Contours} where an acceptable agreement between different methods and the remote-sensing data set is observed (see Table \ref{RMSE_Table}). {We also note that the metrics obtained using the optimized LSTM architecture in this table outperformed the accuracy obtained from classical linear or decision-tree based forecasting techniques (see Table 2 in \cite{maulik2020recurrent})}. A closer examination of the mean squared error during the entire testing period, shown in Figure \ref{NOAA_Contour_RMSE} where POD-RKHS is seen to give sufficiently accurate predictions in the entire domain, including the vital Eastern Pacific region. With the testing period being around 18 years, the El-Ni\~{n}o Southern Oscillation (ENSO) oscillation, occurring at a frequency of 2-7 years \cite{schmidt2019spectral}, is captured more accurately in comparison to CESM.  

In terms of times to solution, data-driven models provided instantaneous forecasts for the given time period (1981--2018). In contrast, equation-based models would require larger computing resources and times-to-solution by several orders of magnitude. For example, CESM (for the forecast period of 1920--2100), required 17 million core-hours on Yellowstone, the National Center for Atmospheric Research (NCAR) high-performance computing resource, for obtaining forecasts for each member of a 30 member ensemble. While finer details about computational costs were not readily available for HYCOM, this short-term ocean prediction system runs daily at the Navy Department of Defense Supercomputing Resource Center, with data typically accessible within 48 hours of the simulation initialization\footnote{https://www.hycom.org/dataserver}. Benchmarking results for the 1/25 degree HYCOM forecasts (much finer than the reference data used here) indicate the requirement of 800 core hours per day of forecast on a Cray XC40 system\footnote{https://www.hycom.org/attachments/066\_talk\_COAPS\_17a.pdf}. For a more appropriate comparison, we also take into account the computational cost for the search of an optimal LSTM architecture and its training. While the POD-RKHS method requires approximately 40 seconds to train on a single node machine without acceleration, the LSTM being compared with was discovered using 3 hours of wall time on 128 compute nodes of the Theta supercomputer with the Intel Knights Landing architecture. Note that POD-RKHS was trained only once on one Intel CoreI7 x86\_64 CPU without any hardware acceleration. Subsequent assessments on a different data set (in the next set of experiments) indicate that the RKHS continues to be competitive in training costs. {All experiments involving POD-RKHS were performed using Python with the NumPy and AutoGrad libraries being used for numerical linear algebra and optimization, respectively.}

\begin{table}
\centering
\caption{RMSEs (in Celsius) for different forecast techniques compared against the POD-RKHS forecasts between April 5, 2015, and June 24, 2018, in the Eastern Pacific region (within -10 to +10 degrees latitude and 200 to 250 degrees longitude). POD-RKHS matches the accuracy of the process-based models for this particular metric and assessment and that of an optimized LSTM \cite{maulik2020recurrent} using neural architecture search.}
\begin{tabular}{|c|c|c|c|c|c|c|c|c|}
\hline
\multicolumn{1}{|c|}{} & \multicolumn{8}{c|}{RMSE ($^\circ$Celsius) }\\
\hline
 & Week 1 & Week 2 & Week 3 & Week 4 & Week 5 & Week 6 & Week 7 & Week 8 \\ \hline
POD-LSTM & 0.62 & 0.63 & 0.64 & 0.66 & 0.63 & 0.66 & 0.69 & 0.65  \\ \hline
CESM      & 1.88 & 1.87 & 1.83 & 1.85 & 1.86 & 1.87 & 1.86 & 1.83  \\ \hline
HYCOM     & 0.99 & 0.99 & 1.03 & 1.04 & 1.02 & 1.05 & 1.03 & 1.05 \\ \hline
POD-RKHS  & 0.76 & 0.67 & 0.66 & 0.69 & 0.69 & 0.72 & 0.77 & 0.76 \\ \hline
\end{tabular}
\label{RMSE_Table}
\end{table}

\begin{figure}
    \centering
    \mbox{
    \subfigure[Coefficient 1]{\includegraphics[width=0.3\textwidth]{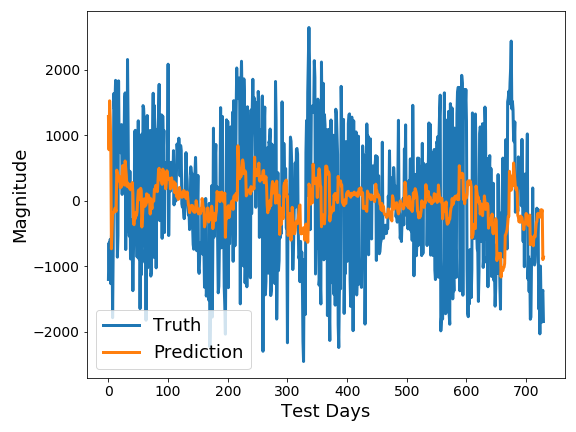}}
    \subfigure[Coefficient 2]{\includegraphics[width=0.3\textwidth]{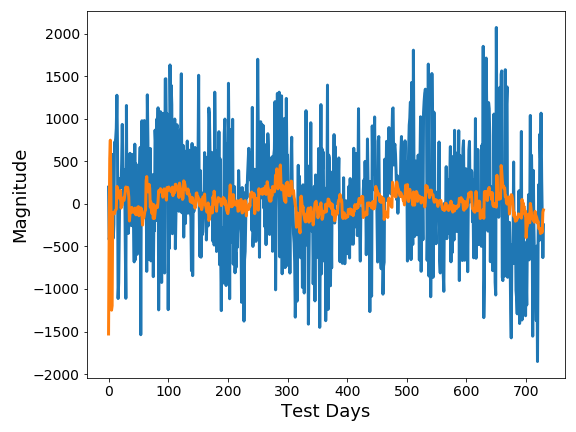}}
    \subfigure[Coefficient 3]{\includegraphics[width=0.3\textwidth]{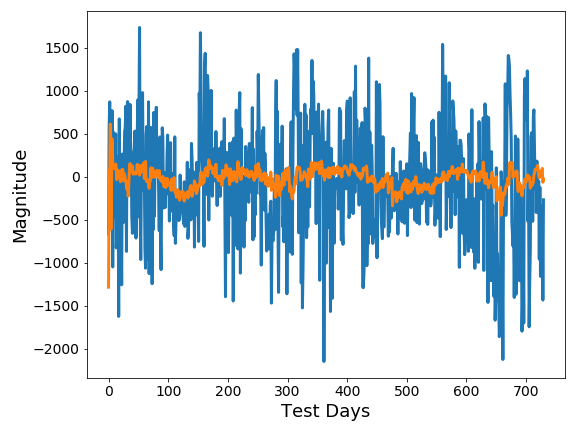}}
    }
    \caption{POD coefficient predictions for forecasting in the testing regime for the NOAA-NCEP NAM data set. The plots indicate the stochastic nature of the daily temperature at midnight and also highlight the ability of the POD-RKHS framework to extract a signal from them.}
    \label{NAM_Coefficients}
\end{figure}

Our second set of assessments are performed for the NOAA-NCEP NAM data set for sea and land surface air temperatures over the North-American continent. This data set is obtained by a re-analysis of numerical simulation data obtained through data assimilation of observations. We follow the same strategy of time-delayed inputs to outputs for forecasting the midnight temperature. Figure \ref{NAM_Coefficients} shows the ability of the time-series emulator to extract an underlying trend from a highly noisy signal.

\begin{figure}
    \centering
    \mbox{
    \subfigure[POD-RKHS Prediction]{\includegraphics[width=0.45\textwidth]{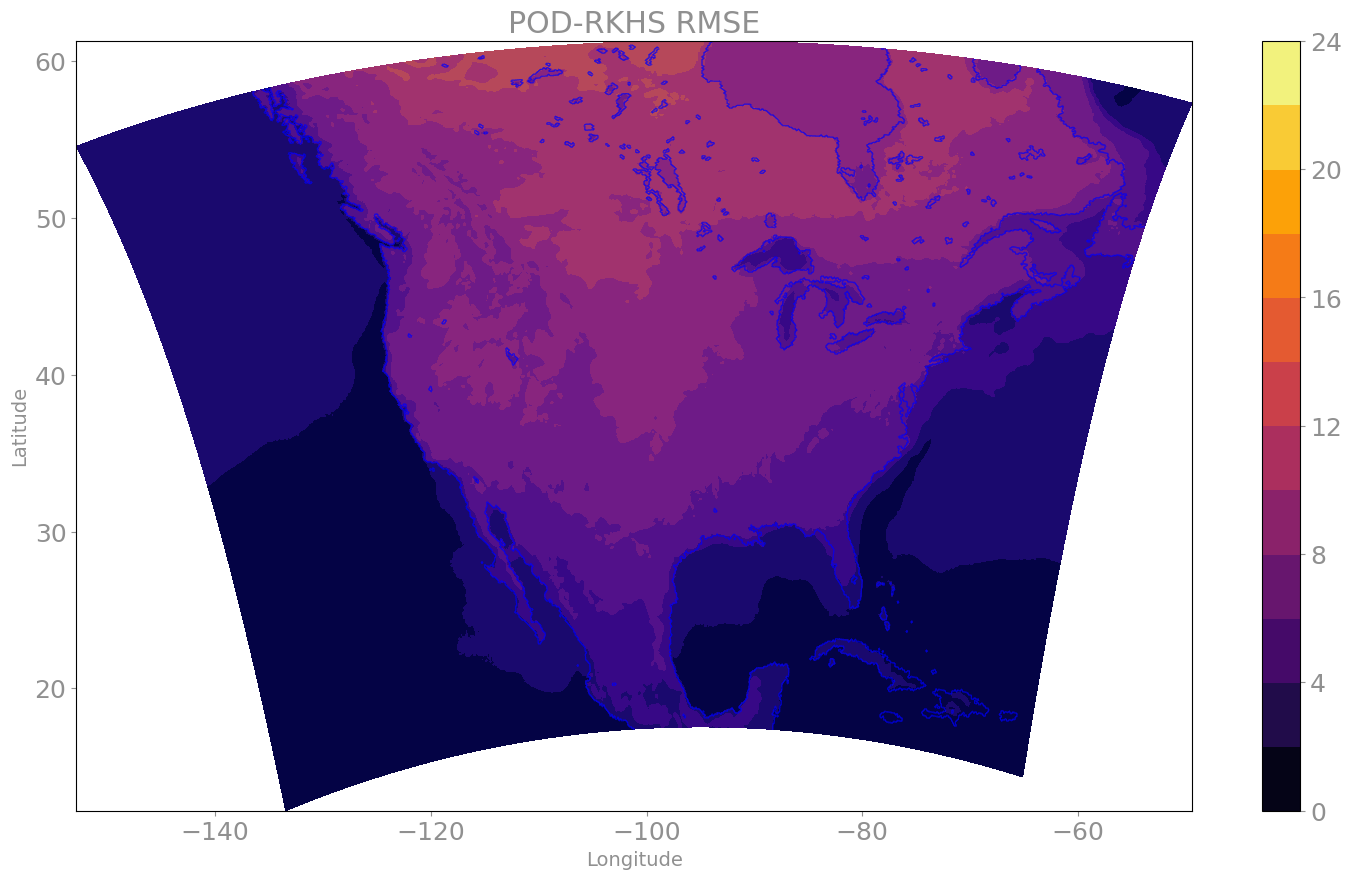}}
    \subfigure[POD-LSTM]{\includegraphics[width=0.45\textwidth]{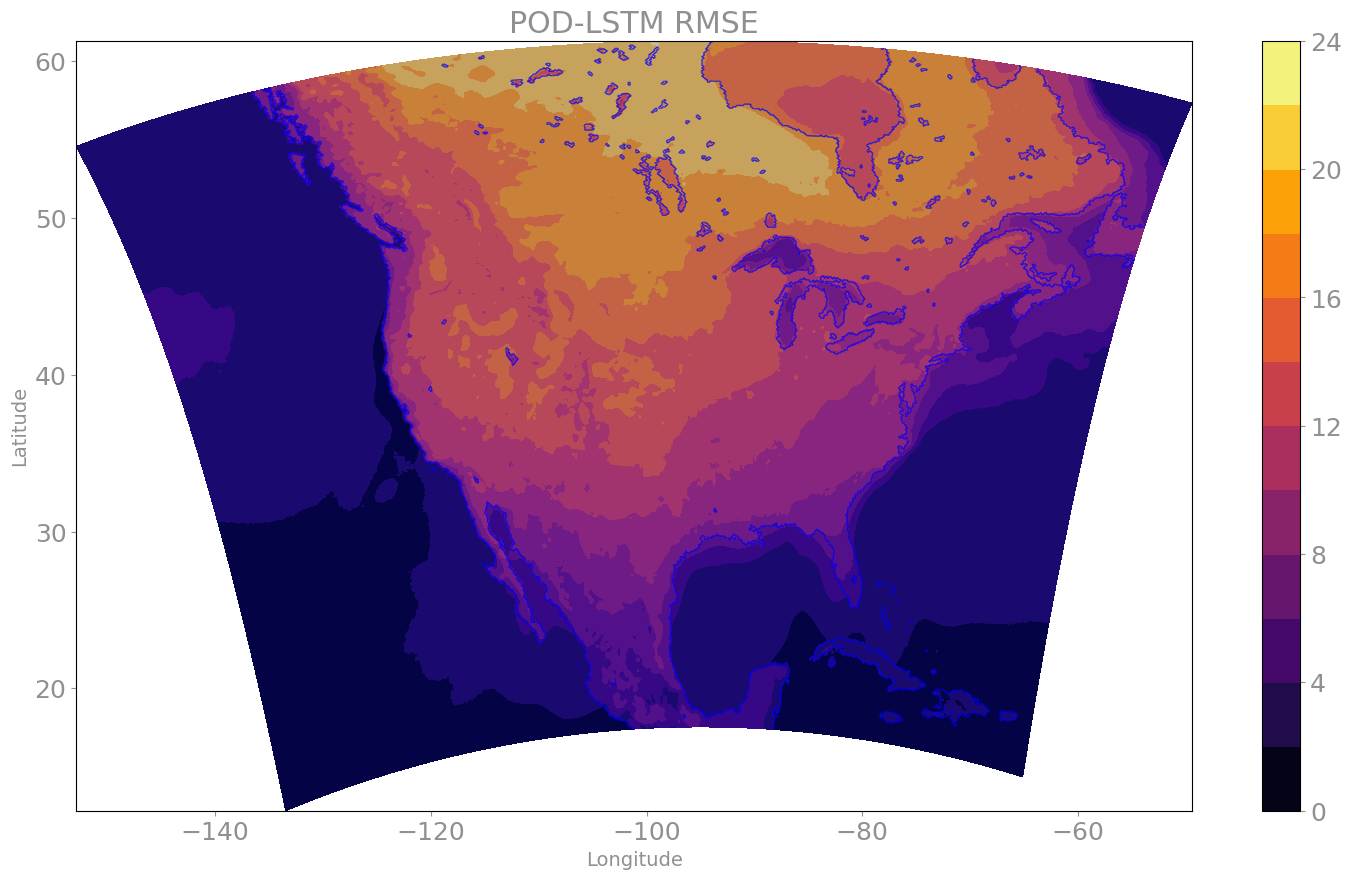}}
    } \\
    \mbox{
    \subfigure[Climatology]{\includegraphics[width=0.45\textwidth]{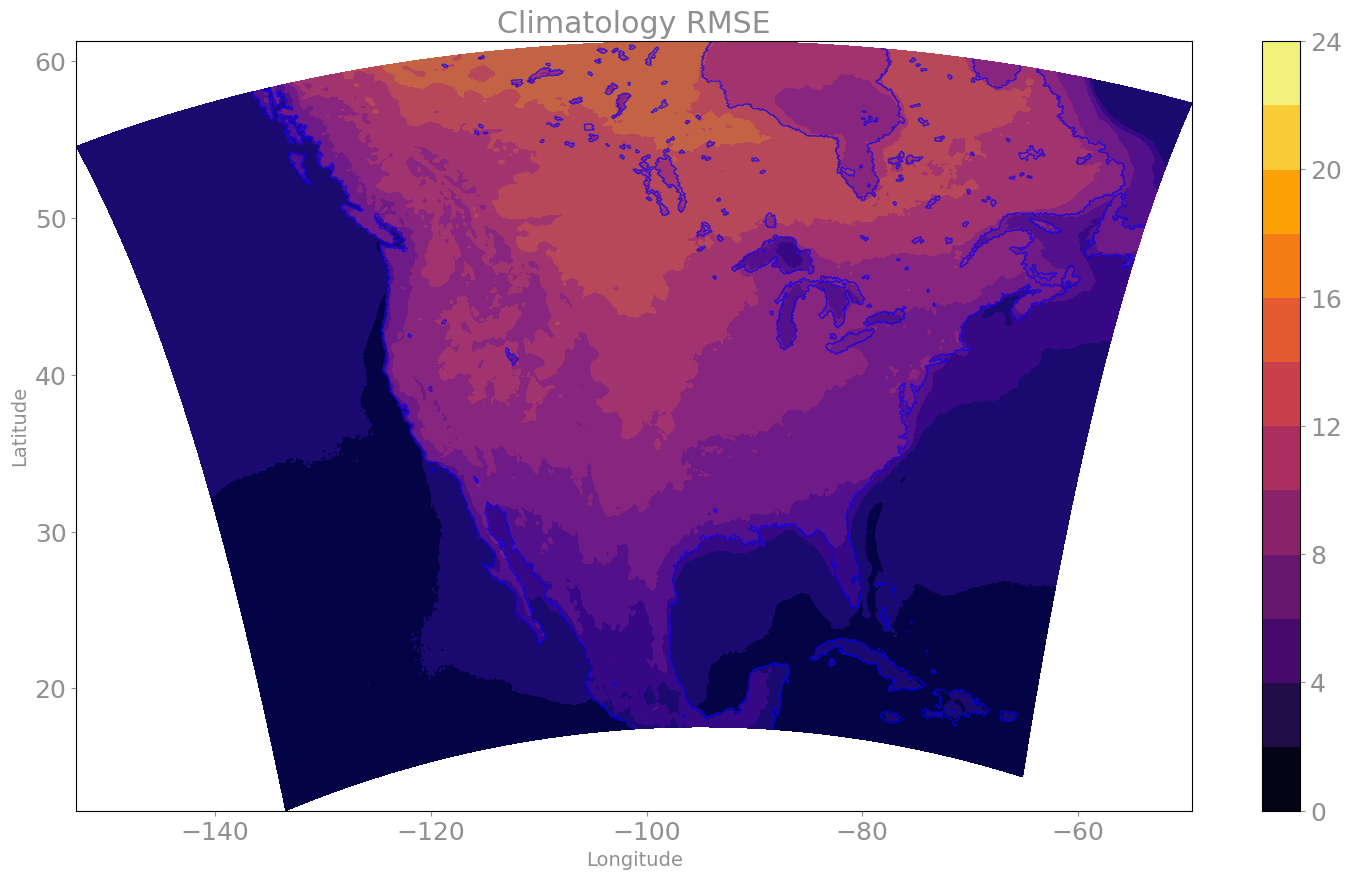}}
    \subfigure[Persistence]{\includegraphics[width=0.45\textwidth]{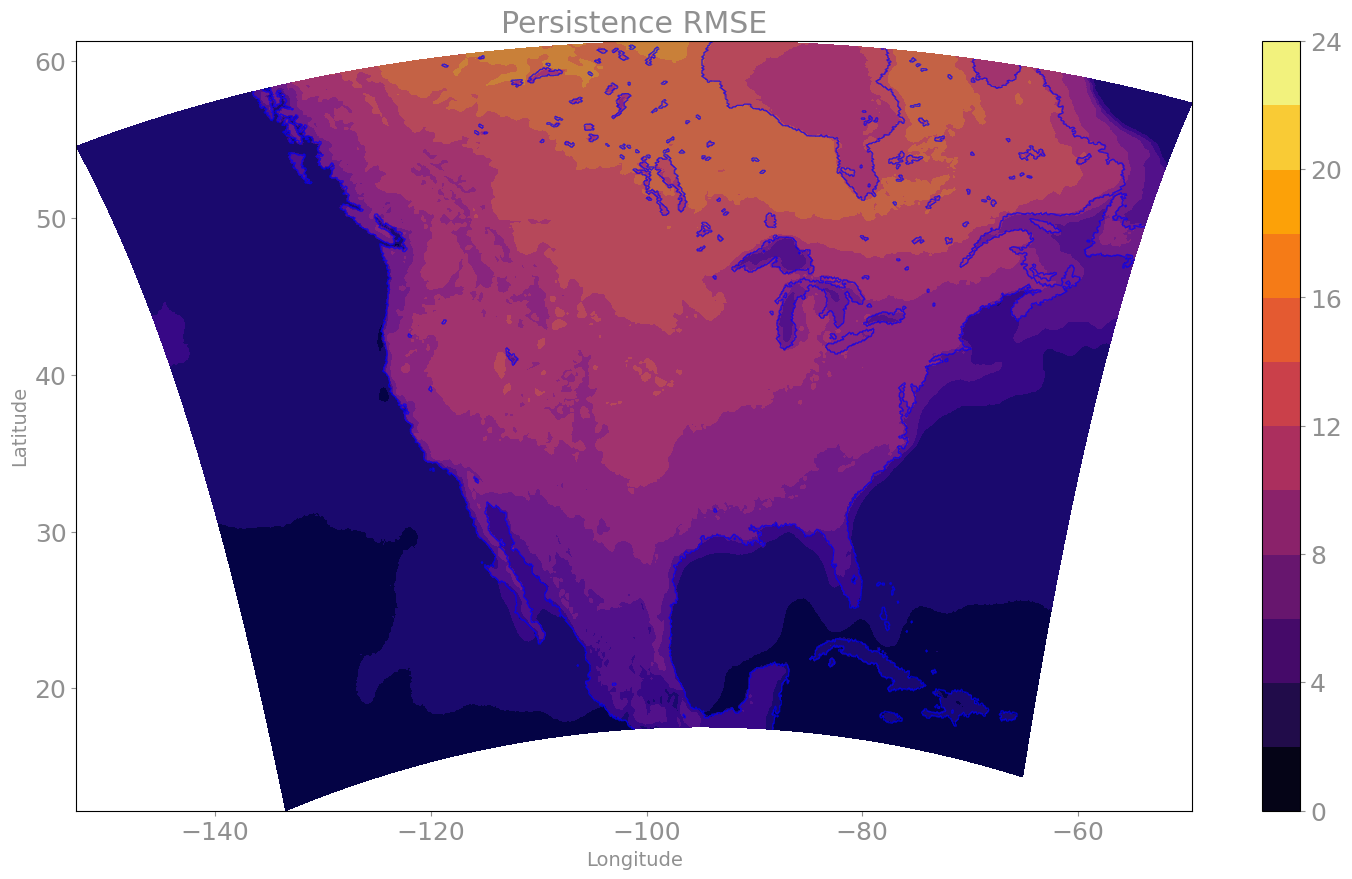}}
    } \\
    \caption{Testing RMSEs for predictions on the NAM data as compared to climatology and persistence baselines for the POD-RKHS and POD-LSTM frameworks. This result indicates the use of one potential LSTM with manual tuning - further improvements may be obtained through neural architecture and hyperparameter search as seen in \cite{maulik2020recurrent}.}
    \label{NAM_1}
\end{figure}

\begin{figure}
    \centering
    \mbox{
    \subfigure[Correlation coefficients]{\includegraphics[width=0.45\textwidth]{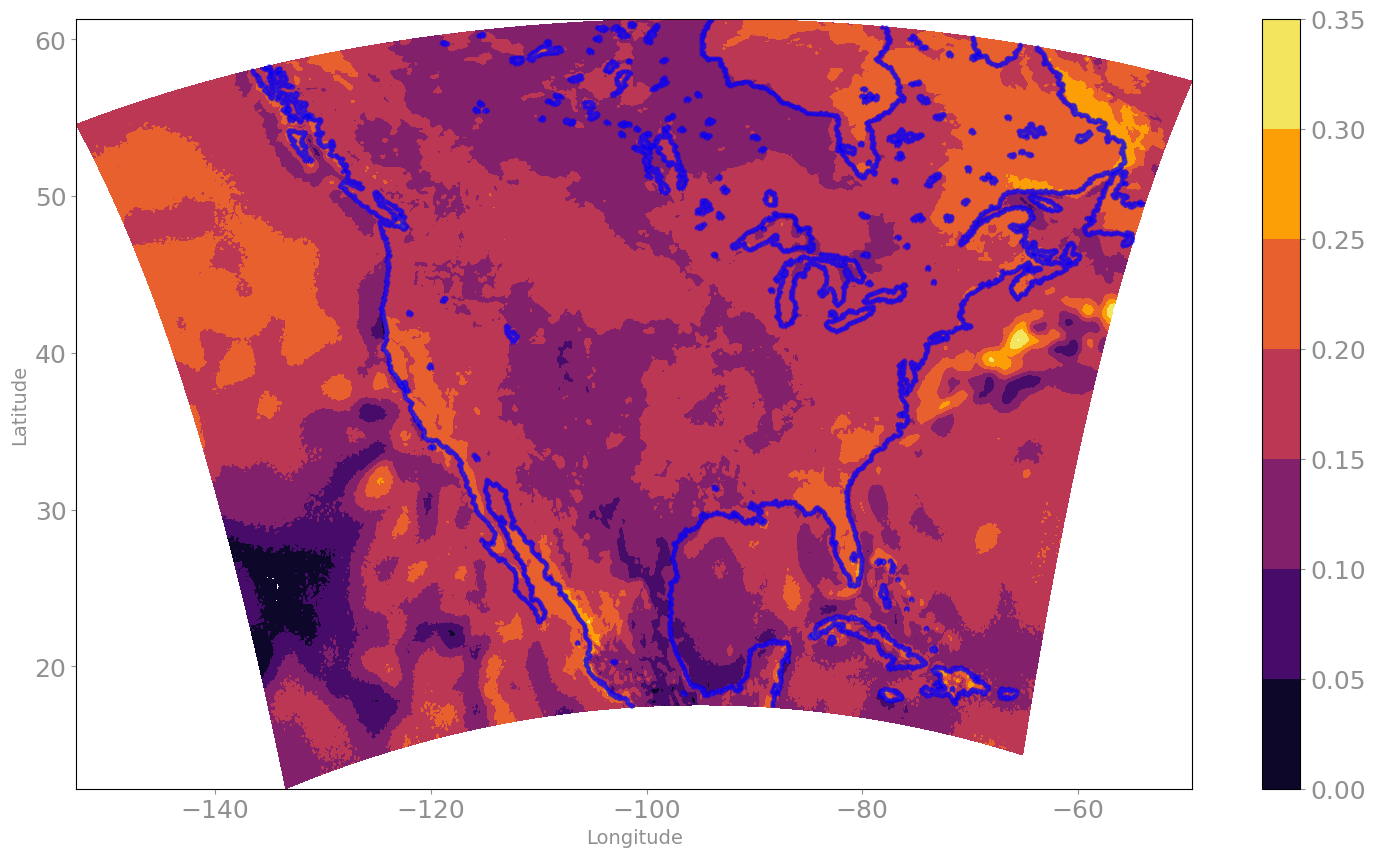}}
    \subfigure[Cosine similarities]{\includegraphics[width=0.45\textwidth]{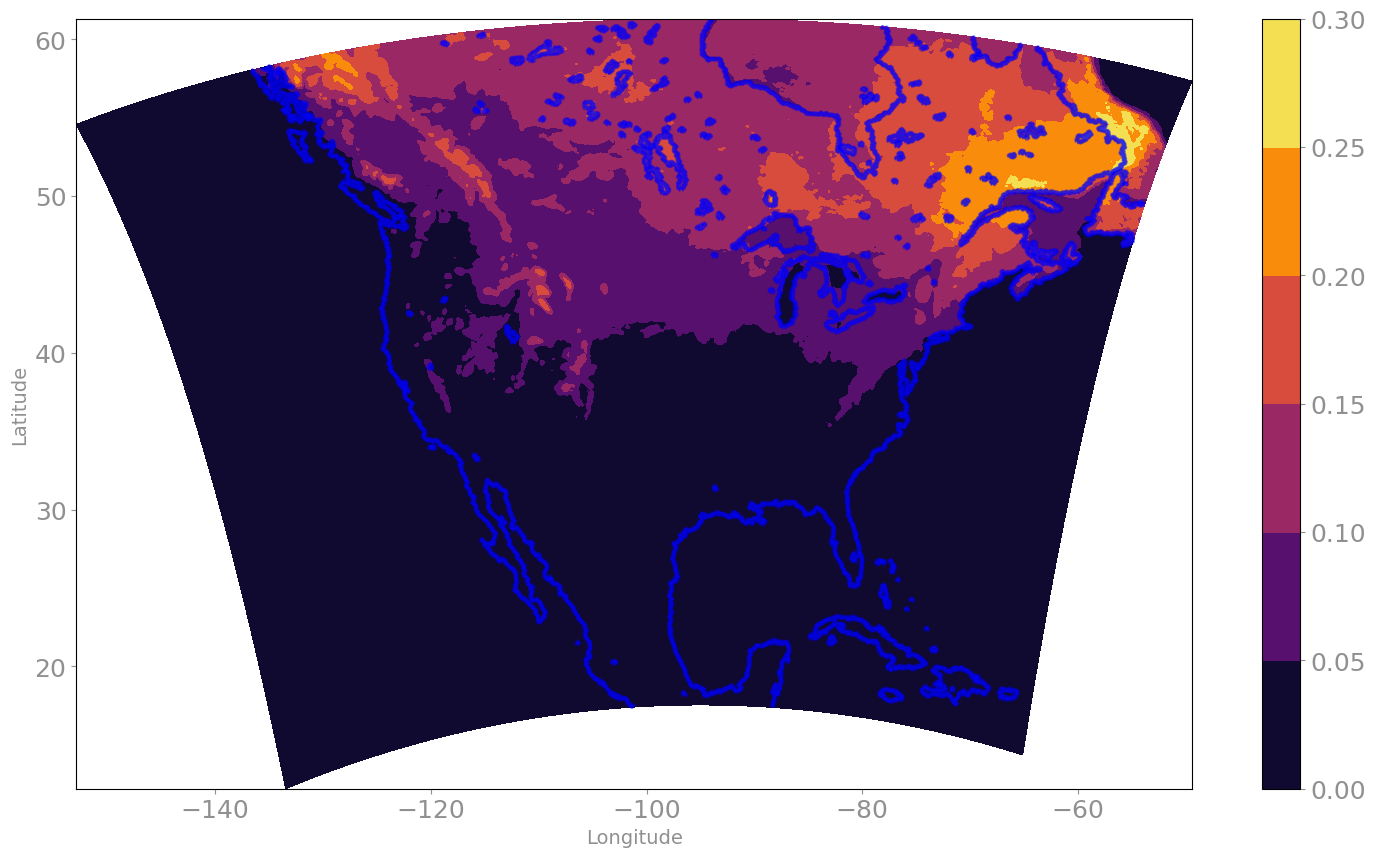}}
    } \\
    \mbox{
    \subfigure[Correlation coefficients]{\includegraphics[width=0.45\textwidth]{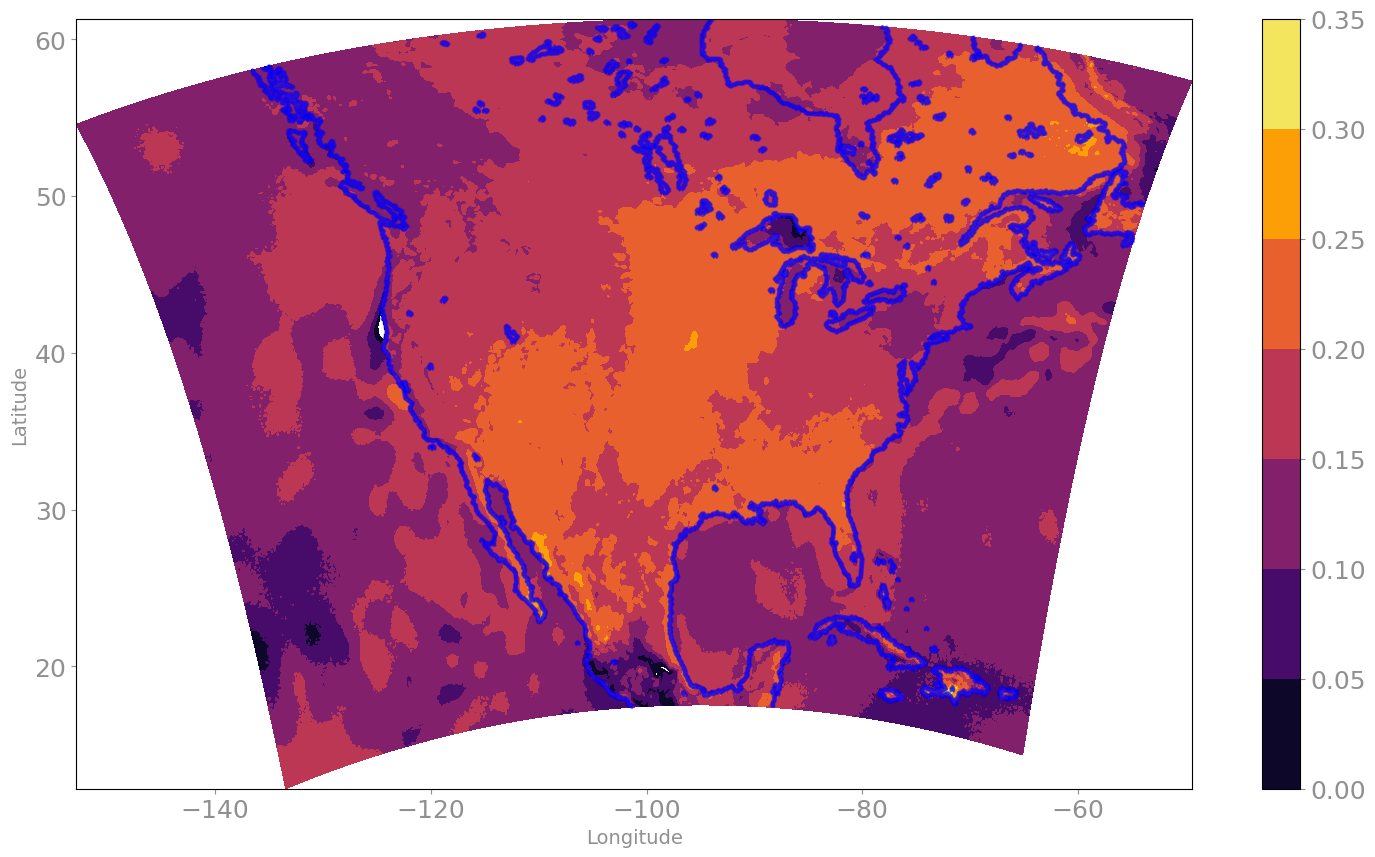}}
    \subfigure[Cosine similarities]{\includegraphics[width=0.45\textwidth]{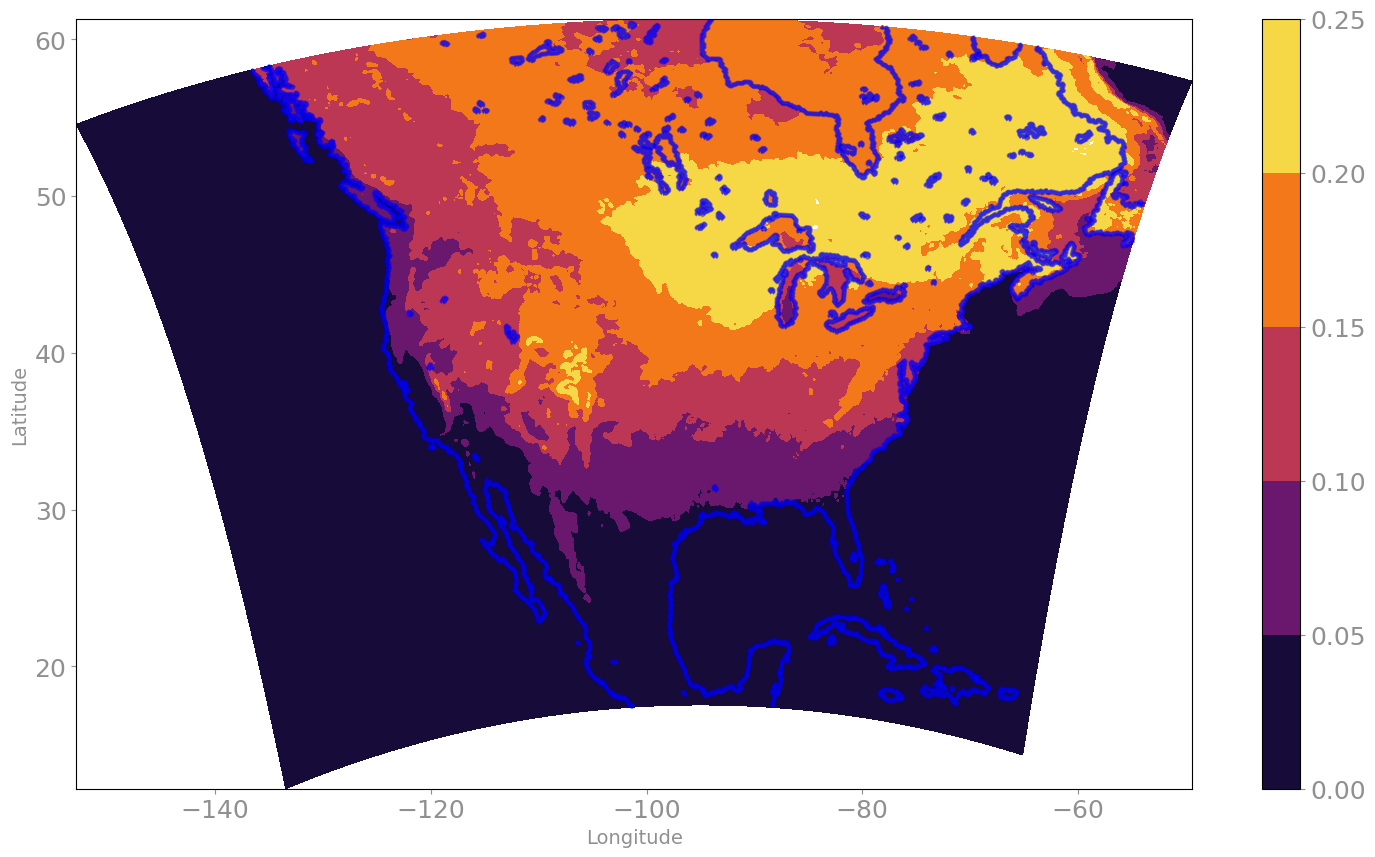}}
    } 
    \caption{Metric \textbf{improvements} on climatology (top) and persistence (bottom) using the POD-RKHS time-series emulator. {Lighter regions depict areas where the POD-LSTM forecast model outperformed the climatology and persistence baselines.} The north domain of the data set shows improved performance by the proposed framework, particularly for extreme fluctuation detection (indicated by Cosine similarity improvement). Overall correlation with the truth is seen to be improved throughout the domain.}
    \label{NAM_2}
\end{figure}

Note that the difficulty of extracting the extremely high-frequency features contributes to the poor capture of extreme events during emulation. The reconstruction accuracy from the forecast is compared in a series of assessments beginning with root-mean-squared error (RMSE) assessments as shown in Figure \ref{NAM_1} for the testing time period. POD-RKHS is seen to provide competitive results in comparison to persistence and climatology at the lower latitudes. {Here, climatology refers to the temperature on a particular day averaged over multiple years, and persistence is the last-known (i.e., last observed) temperature on the day the forecast is made.} In contrast, exceptional gains are seen at the higher latitudes with far reduced RMSE values from the proposed framework. Note that errors across all frameworks were maximum in the Northern section of the continent, proving the increasing difficulty of predicting in that region using data-driven techniques. Figure \ref{NAM_2} plots the contours for improvements in the correlation coefficient and cosine similarities \footnote{The cosine similarity between two vectors $\mathbf{a}$ and $\mathbf{b}$ is $\frac{\mathbf{a} \cdot \mathbf{b}}{||\mathbf{a}|| ||\mathbf{b}||}$} {of the predictions when compared to climatology}. While correlation coefficients indicate regions where the predictions improve over the baselines in an averaged sense, Cosine similarities indicate the ability to detect extreme fluctuations in a forecast. The contours indicate that the proposed framework improves on extreme fluctuation detection over climatology in the north region of the spatial domain as well. Similar trends are observed for comparisons with persistence with large gains in Cosine similarities in the Northern part of the spatial domain. 

\begin{figure}
    \centering
    \mbox{
    \subfigure[Coefficient 1]{\includegraphics[width=0.3\textwidth]{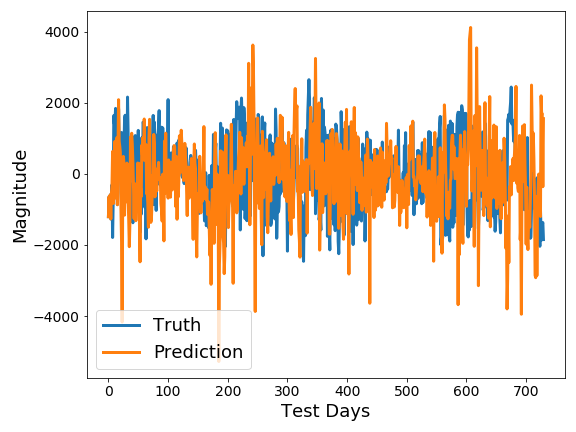}}
    \subfigure[Coefficient 2]{\includegraphics[width=0.3\textwidth]{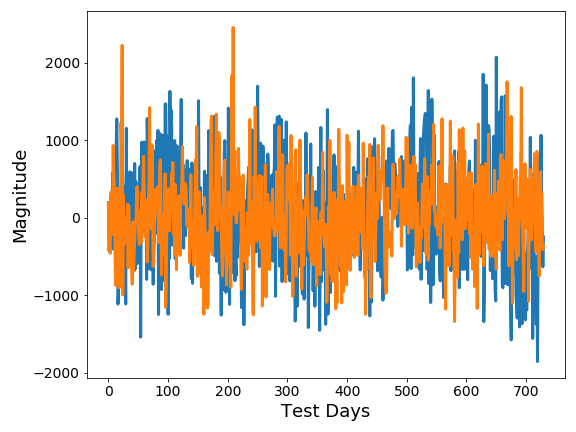}}
    \subfigure[Coefficient 3]{\includegraphics[width=0.3\textwidth]{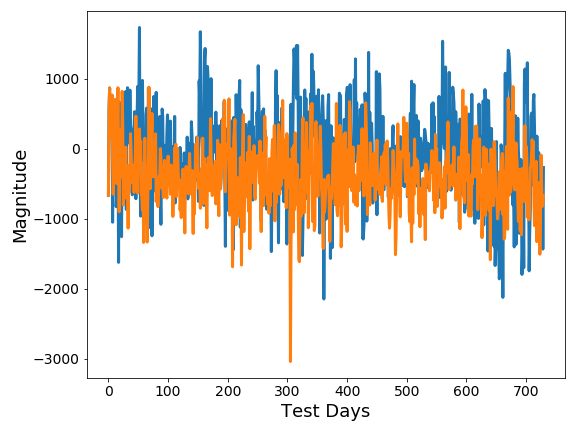}}
    }
    \caption{Predictions for the NOAA-NCEP NAM POD Coefficients using a POD-LSTM framework. This prediction was obtained with 3 stacked LSTM cells, each with 50 neurons for all the internal operations. An LSTM that provided the best training and validation performance among 55 restarts was chosen. The total time to train these LSTMs was 5 hours on an Nvidia V100 GPU. It is observed that the LSTM architecture is sensitive to the stochastic nature of the data.}
    \label{LSTM_Coefficients}
\end{figure}

\begin{figure}
    \centering
    \mbox{
    \subfigure[Correlation coefficients]{\includegraphics[width=0.45\textwidth]{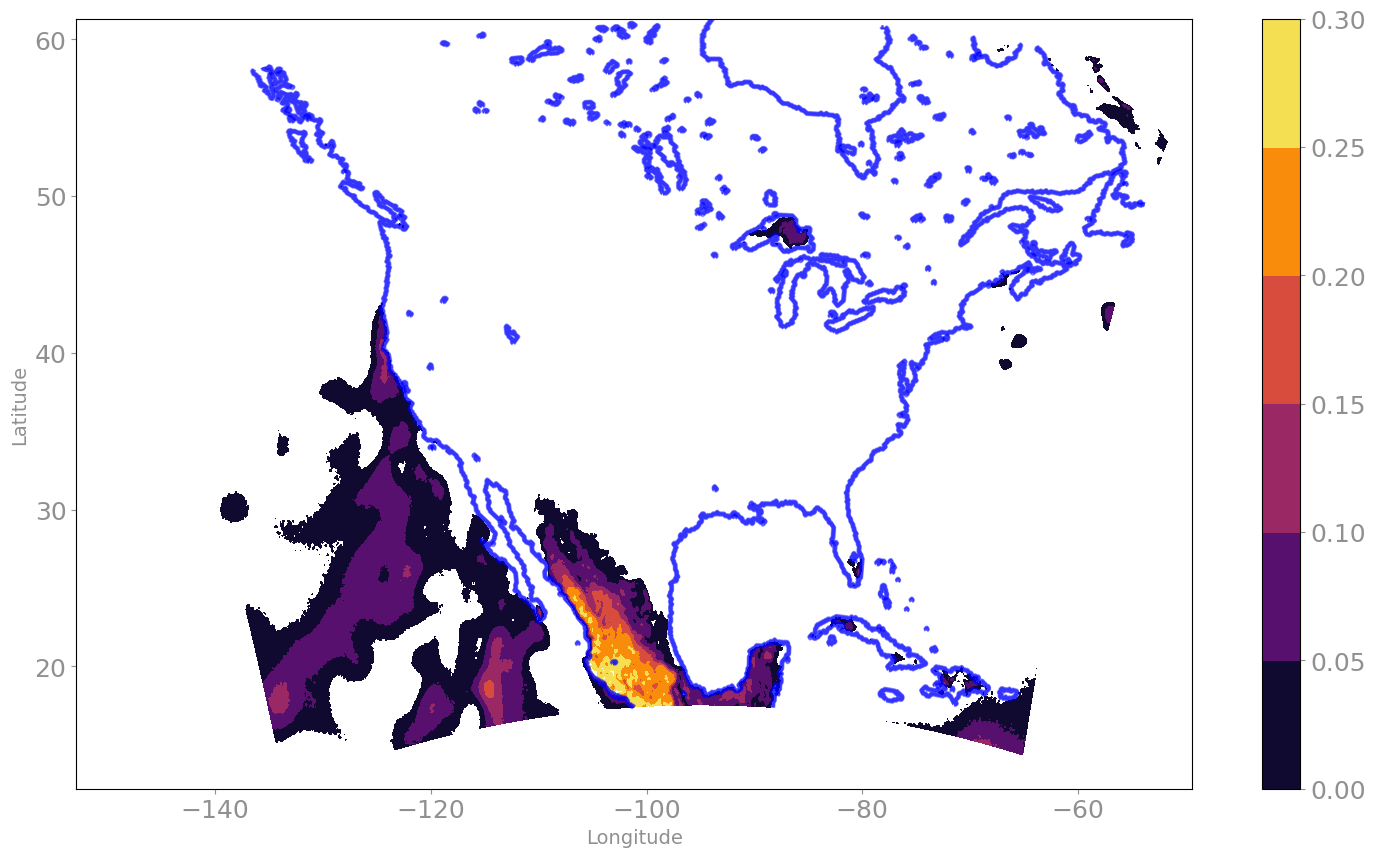}}
    \subfigure[Cosine similarities]{\includegraphics[width=0.45\textwidth]{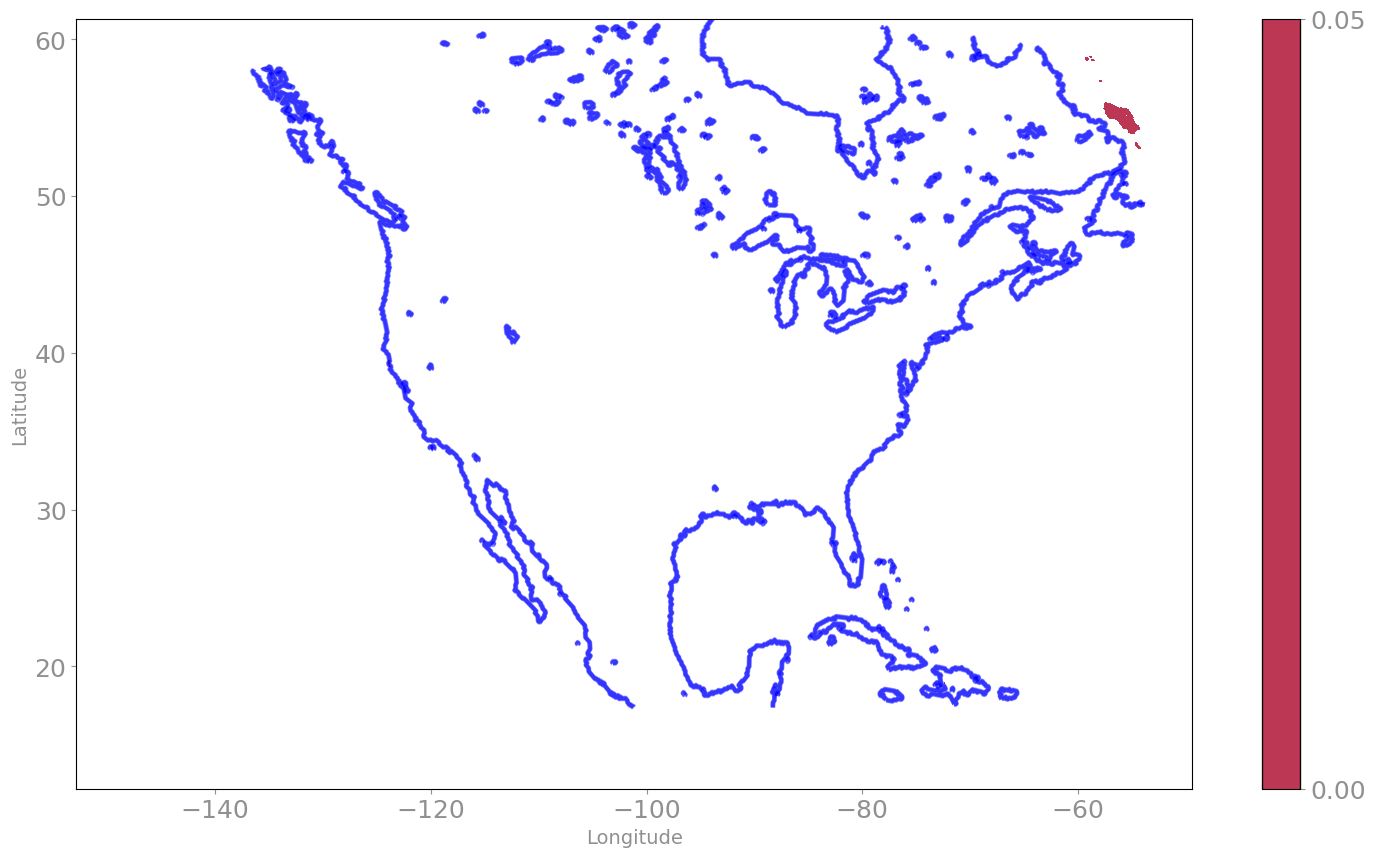}}
    } \\
    \mbox{
    \subfigure[Correlation coefficients]{\includegraphics[width=0.45\textwidth]{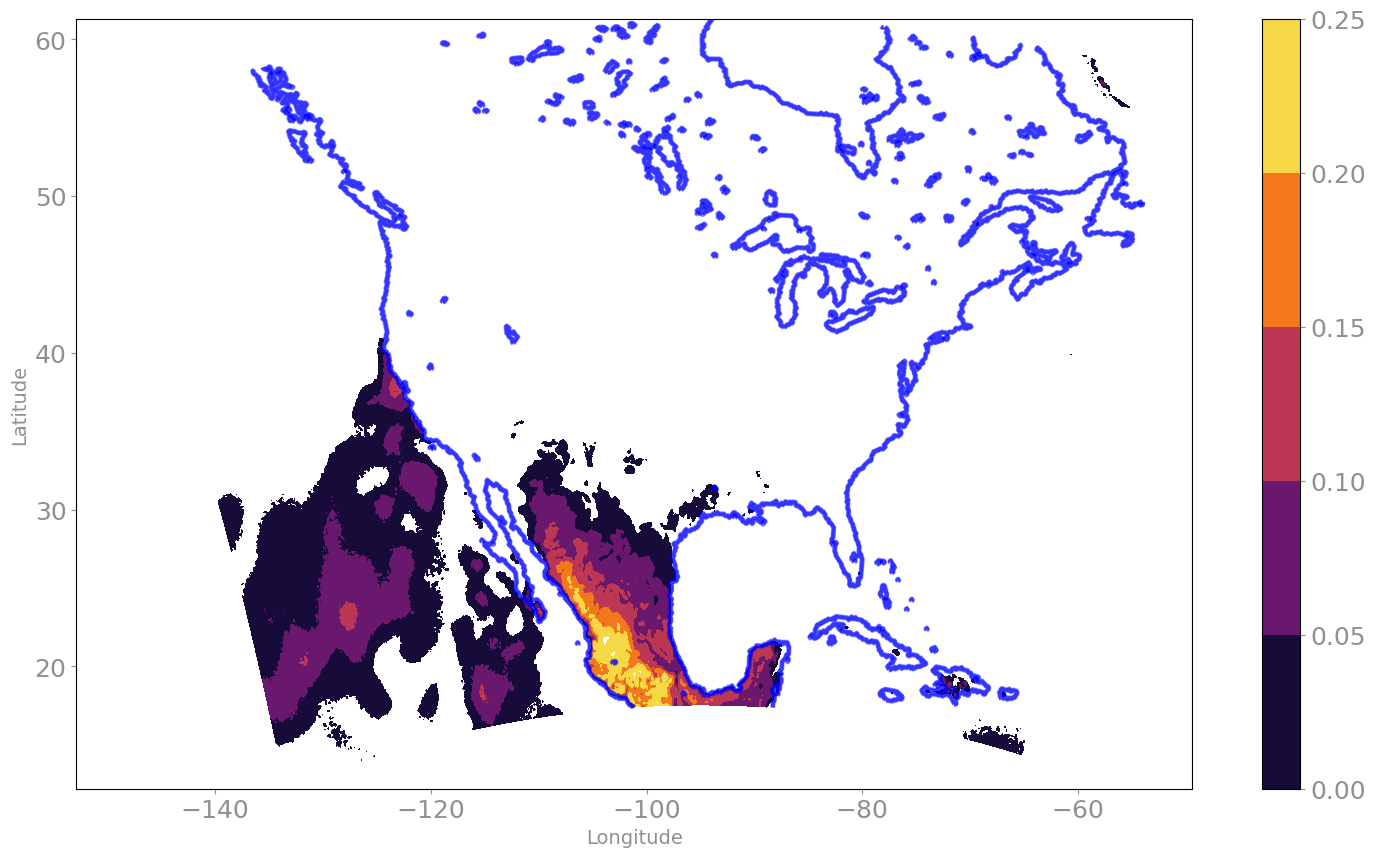}}
    \subfigure[Cosine similarities]{\includegraphics[width=0.45\textwidth]{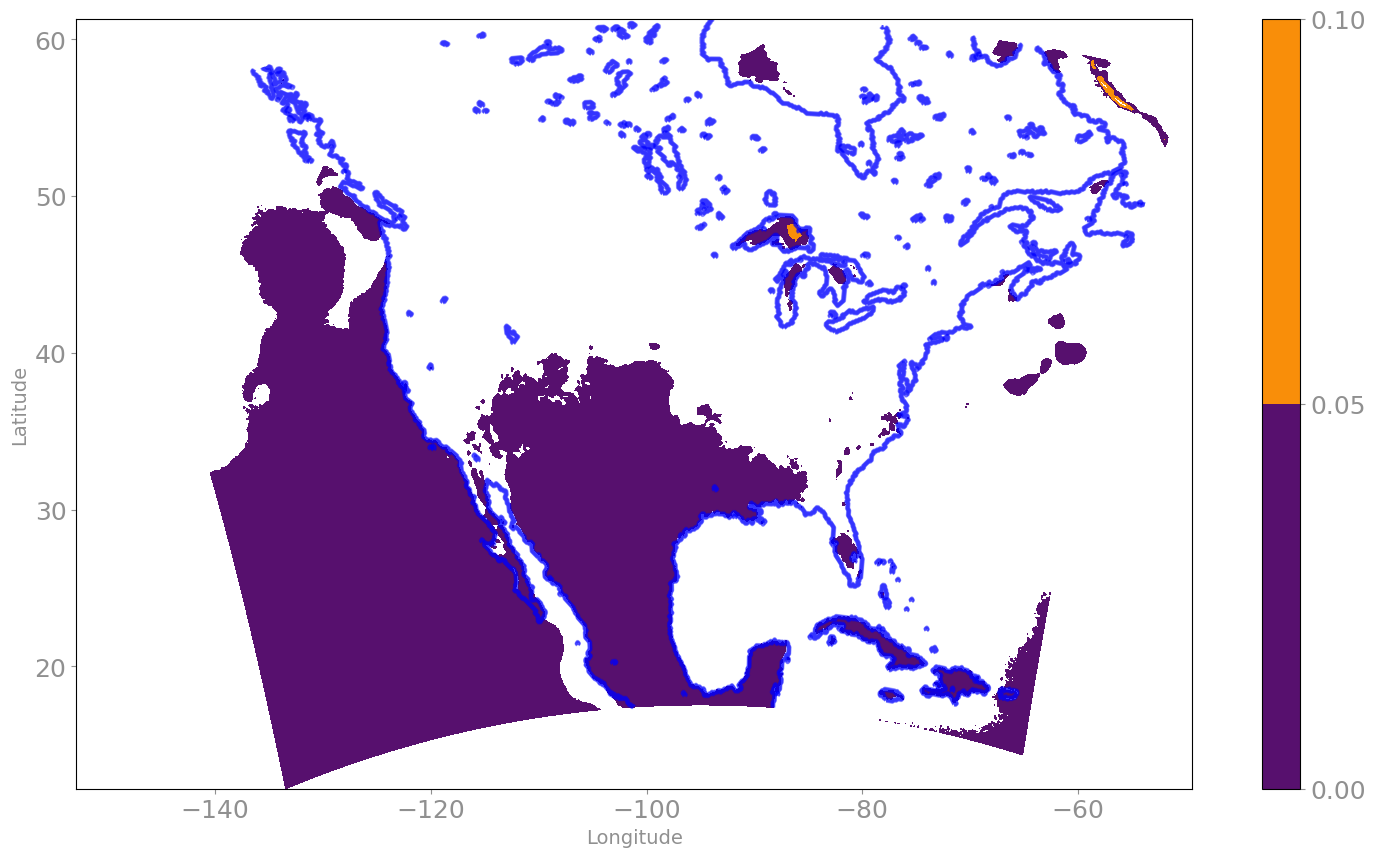}}
    } 
    \caption{Metric \textbf{improvements} on climatology (top) and persistence (bottom) using the LSTM time-series emulator. {Lighter regions depict areas where the POD-LSTM forecast model outperformed the climatology and persistence baselines.} Vast stretches (i.e., the white regions) of the domain show poorer performance for POD-LSTM against climatology and persistence baselines.}
    \label{NAM_3}
\end{figure}

For the purpose of comparison, we also outline the performance of a standard LSTM framework (manually selected in comparison to \cite{maulik2020recurrent}) for forecasting on the NAM data set. The problem formulation is identical, with the LSTM utilizing a window of inputs of 7 days length and forecasting the POD coefficient trajectories for the next 7 days. We train 55 instances of this LSTM, consisting of 3 stacked LSTMs and 50 neurons for each of the linear operations in each cell, and select the one with the best training and validation performance. The 55 LSTM architectures were trained using an Nvidia V100 GPU and required 5 hours of wall time. For each training, an early stopping criterion is utilized to terminate training if validation losses do not improve for ten successive training epochs. Figure \ref{LSTM_Coefficients} shows predictions from the testing data using the trained LSTM with clear indications of poor capture of dynamics. These are confirmed by contour plots for the RMSE, shown in Figure \ref{NAM_1}, where the POD-LSTM framework is the weakest of all the data-driven methodologies presented here. Figure \ref{NAM_3} examines potential improvements in the correlations and Cosine similarities within the spatial domain of the data set for this experiment. Climatology and persistence are seen to comprehensively outperform a standard LSTM architecture. Note that, as suggested by \textcolor{black}{Figures} \ref{NOAA_Coefficients} and \ref{NAM_Coefficients}, most of the forecast error is due to bias for the NOAA-SST dataset (low noise dataset) and variance for the NOAA-NCEP NAM dataset (high noise dataset). It must be noted that this result is not meant to justify mistrust of all deep learning architectures but propose the use of classical kernel-based methods for the construction of baselines.

We now compare the training and testing times of the proposed framework with those of LSTM methods with standard architectures. The POD-RKHS method required a total of 136 seconds to train, whereas the POD-LSTM methods took an average of 330 seconds to train per LSTM architecture. Although the choice of user-defined hyperparameters in both these methods may change these values, their training times are comparable. We note that, in a manner similar to \cite{maulik2020recurrent}, discovering the optimal hyperparameters and architectures for a deep learning framework adds significant costs associated with sampling a high dimensional search space of neural networks that may fit the given training data. In that regard, kernel-methods may provide significant computational gains. 

\section{Discussion}

In this article, we have employed a simple {kernel  regression} method for the forecasting of geophysical time series.  We have shown that when the kernel is also learned from data (via Kernel Flows),  then the proposed framework provides competitive and computationally efficient baselines for the geophysical emulation of two temperature-based data sets given by the weekly averaged sea-surface temperature (NOAA-SST) and the daily resolved midnight temperature (NCEP-NAM). In both cases, the proposed method recovers a stable temporal behavior on a low-dimensional manifold. Importantly, for the second data set, strong stochasticity is handled robustly by the proposed framework to obtain superior metrics when compared to classical baselines such as climatology or persistence. Comparisons with the POD-LSTM framework, commonly used for non-intrusive reduced-order modeling, are also favorable. While the POD-LSTM obtained through expensive parallelized neural architecture search slightly outperforms the proposed framework for the relatively smooth NOAA-SST time-series data, the proposed framework is much more successful in the case of the stochastic NCEP-NAM temperature data. The computational cost of learning kernels in the proposed method is considerably lower than PDE-based forecasting alternatives and does not require any specialized hardware (such as artificial intelligence accelerators). When considering the fact that deep learning frameworks oftentimes require significant architecture tuning to obtain a viable function approximation, our framework provides greater computational gains when compared to those data-driven surrogates over the lifetime of a forecast campaign. 

Extensions to this work include the addition of more predictive variables to the training data to emulate larger fractions of the earth system model. This may be obtained by concatenating reduced representations of a greater number of flow field variables (in truncated POD space) to obtain a larger, more informative state vector. {We note that, despite the successes of the framework in this article, several forecast tasks may be challenging using the POD-RKHS methodology. These would generally be due to the absence of a low-dimensional manifold on which a large portion of the flow-field dynamics is restricted. An example forecast task would be something akin to the prediction of precipitation. Therefore future extensions must account for situations where a linear compression, given by the POD, is inefficient due to the presence of very fine scales in the data - in such cases, nonlinear compression techniques such as diffusion maps or neural network parameterized autoencoders may be necessary for effective dimensionality reduction. Here, an interesting avenue for exploration is to use kernel flows for obtaining reduced representations themselves in a unified pipeline for geophysical emulation.} These directions may be instrumental for pushing data-driven forecast horizons to regimes where classical methods are untrustworthy. {Finally, a major advantage of using kernel flows is that posterior distributions may be learned under the Gaussian assumption, which would allow for aleatoric uncertainty quantification. A further assumption of hyperparameters being random variables that are sampled from a joint distribution would enable epistemic uncertainty estimates as well. These ideas are our current research focus.}

\dataccess{The data that support the findings of this study are openly available in Github at \url{https://github.com/Romit-Maulik/POD\_RKHS}.}

\aucontribute{B.H. prepared code and analysis, wrote portions of the paper. R.M. designed the investigation, prepared code and documentation, performed analyses, generated visualizations, wrote paper. H.O. prepared code and analysis, wrote portions of the paper.}

\ack{This material is based upon work supported by the U.S. Department of Energy (DOE), Office of Science, Office of Advanced Scientific Computing Research, under Contract DE-AC02-06CH11357. This research was funded in part and used resources of the Argonne Leadership Computing Facility, which is a DOE Office of Science User Facility supported under Contract DE-AC02-06CH11357. R.~M. acknowledges support from the Margaret Butler Fellowship at the Argonne Leadership Computing Facility. B.~H. thanks the European Commission for funding through the Marie Curie fellowship STALDYS-792919 (Statistical Learning for Dynamical Systems). H.~O. gratefully acknowledges support by  the Air Force Office of Scientific Research under award number FA9550-18-1-0271 (Games for Computation and Learning) and  MURI (FA9550-20-1-0358). This paper describes objective technical results and analysis. Any subjective views or opinions that might be expressed in the paper do not necessarily represent the views of the U.S. DOE or the United States Government.}

\bibliographystyle{plain}
\bibliography{references}

\section*{Appendix}

\subsection*{A. Gaussian process regression}
 The interpolant \eqref{mean_gp} can also be identified as the  conditional mean of the centered GP $\xi\sim \mathcal{N}(0,K)$ with $K$ as the covariance function conditioned on $\xi(X_i)=Y_i$, 
 Furthermore, we have the pointwise error estimate  (see \cite{wu1993local} and \cite[Thm.~5.1]{Owhadi:2014}) 
 \begin{equation}\label{eqkejbdkddbs}
|f^\dagger(x)-f(x)|\leq \sigma(x) \|f^\dagger\|_K, \end{equation} 
 bounding interpolation error between $f^\dagger$ and $f$, where {$\|f^\dagger\|_K$ is the RKHS norm defined by the kernel $K$, and}
 \begin{equation}\label{variance_gp}
\sigma^2(x)=K(x,x)-K(x,X) (K(X,X))^{-1} K(x,X)^T\,.
\end{equation}
is the conditional variance of the GP $\xi$. { From a practical point if view $\|f^\dagger\|_K$ could be replaced by the RKHS norm of its interpolant $\sqrt{Y^T (K(X,X))^{-1} Y}$ to offer a rough error estimate}\footnote{{Since $\sqrt{Y^T (K(X,X))^{-1} Y}$ is only a lower bound on $ \|f^\dagger\|_K$, this modified estimate is no longer a rigorous bound. We also note that the effect of KF is to decrease both  $\sqrt{Y^T (K(X,X))^{-1} Y}$ and $ \|f^\dagger\|_K$.}}

\subsection*{B. Error Estimates}
For the sake of completeness, we include convergence results from \cite{BHSIAM2017, lyap_bh} that characterize the error estimates 
of the difference between a dynamical system defined by an ODE and its approximation from data.

We consider ordinary differential equations of the form
\begin{equation}
  \dot{x}  = f^*(x), 	\label{eqn:dynsys}
\end{equation}
where $f^*: \mathbb{R}^d \to \mathbb{R}^d$ is a smooth vector field and dot denotes differentiation with respect to time.

We assume that the function $f^*$ is unknown, but we have sampled data of the form $(x_i, y_i)$ in $X\times \mathbb{R}^d$, $i=1,\dots,m$, with \begin{equation}\label{noisy_measurement} y_i = f^*(x_i) + \eta_{x_i}\end{equation}
We assume that the one-dimensional random variables $\eta_{x_i}^k \in \mathbb{R}^d$, where $i=1,\dots,m$ and $k=1,\dots,d$, are independent random variables drawn from a probability distribution with zero mean and variance $(\sigma_{x_i}^k)^2$ bounded by $\sigma^2$. 

The function spaces that we  use to search for our approximations to both $f^*$  will be  reproducing kernel Hilbert spaces (RKHS). For a survey of the main properties of RKHS spaces mentioned in this section, we refer to \cite{CuckerandSmale}.

Let $K$ be a continuous, symmetric, positive definite function (a ``kernel'') $K:X\times X\rightarrow\mathbb{R}$, and set $K_x:= K(\cdot,x)$. Define the Hilbert space $\mathcal{H}_K$ by first considering all finite linear combinations of functions $K_x$, that is $\sum_{x_i\in X} a_iK_{x_i}$ with finitely many $a_i\in\mathbb{R}$ nonzero. An inner product $\langle \cdot , \cdot \rangle_K$ on this space is defined by $\langle K_{x_i},K_{x_j} \rangle_H := K(x_i,x_j)$ and extending linearly. One takes the completion to obtain $\mathcal{H}_K$.

Alternatively, an equivalent definition of an RKHS is as a Hilbert space of real-valued functions on $X$ for which the evaluation functional $\delta_x(f):= f(x)$ is continuous for all $x\in X$.

Finite dimensional subspaces of $\mathcal{H}_K$ can also be naturally defined by taking a finite number of points $\mathbf{x}:=\{x_1,\ldots,x_m\}\subset X$ and considering the linear span $\mathcal{H}_{K,\mathbf{x}} := \textrm{span}\{K_x:x\in\mathbf{x}\}.$
In practice, we will seek functions in these finite-dimensional subspaces as approximations for $f^*$.

RKHSs are characterized by the reproducing property:
\begin{equation}
\langle K_x,f\rangle_K = f(x), \qquad \forall f\in\mathcal{H}_K.		\label{eq:reproducing}
\end{equation}
If we denote $\kappa:=\sqrt{\sup_{x\in X} K(x,x)}$, then $\mathcal{H}_K\subset C(X)$ and it follows that
\begin{equation}
||f||_{L^\infty(X)} \le \kappa||f||_K, \qquad \forall f\in\mathcal{H}_K.		\label{eq:supnorminclusion}
\end{equation}

The RKHS $\mathcal{H}_K$ can also be defined by means of an integral operator. Let $\rho$ be any (finite) strictly positive Borel measure on $X$ (e.g. Lebesgue measure) and ${L}^2_\rho(X)$ be the Hilbert space of square integrable functions on $X$. Then define the linear operator $L_K:{L}^2_\rho(X)\rightarrow C(X)$ by $ (L_K f)(x) = \int_X K(x,y)f(y)d\rho(y).$ The RKHS will then correspond to the closure of the span of the eigenfunctions of $L_K$.

\begin{theorem} \label{thm_errors} Let $K$ be a Mercer kernel  such  that $K \in C^{2s+\epsilon}(X\times X)$ for $0<\epsilon<2$ and suppose that  $L_K^{-r} f^* \in \mathcal{L}^2_\rho(X)$ for some $0 < r\le 1$. 
 Let $\lambda > 0$, then for every $0 < \delta < 1$, with probability $1 - \delta$, we have
$$\hspace{-4mm}	||f_{\mathbf{z},\lambda} - f^*||_\infty \le 
\frac{||\mathbf{w}||_{\mathbf{R}^m}\sigma\kappa^2}{\lambda \sqrt{\delta}} + \frac{\kappa C \, h_\mathbf{x} \rho(X)}{\lambda} + \kappa\lambda^{r-\frac{1}{2}}||L_K^{-r}f^*||_{\infty}$$
where $\sigma^2:=\sup_{x\in X}\sigma_{x_i}^2$,  $\kappa:=\sqrt{\sup_{x\in X} K(x,x)}$, and $C$ is a Lipschitz constant for $(f^* - f_\lambda)$, where the constant $C$ depends on $f^*$,  $d$, $\sigma$ and the choice of RKHS $\mathcal{H}_{K^1}$.
\end{theorem}
 
The three terms in theorem \ref{thm_errors} correspond, respectively, to errors incurred by the \emph{noise} (\emph{sample error}, ${\mathcal E_1}$), the \emph{finite set of data sites} (\emph{integration error},  ${\mathcal E_2}$ ) and the \emph{regularisation parameter} $\lambda$ ({\emph{regularisation error}, ${\mathcal E_3}$}).

\end{document}